\documentclass[english,pra,superscriptaddress,reprint,twocolumn]{revtex4-1}
\usepackage{standalone}
\usepackage{amsmath}
\usepackage{bbold}
\usepackage{nccfoots}
\usepackage{comment}
\usepackage{graphicx}
\graphicspath{{Figures/}}
\usepackage{braket}
\usepackage{color}
\usepackage{nicefrac}
\usepackage{braket}
\usepackage[utf8]{inputenc}
\usepackage{chngcntr}

\usepackage[colorlinks=true,
   linkcolor=blue,
   citecolor=blue,
   urlcolor =blue]{hyperref}

\newcommand{\citeSM}{\cite[{\tiny SM}\kern-0.3em][]{SM}}

\newcommand{\be}{\begin{equation}}
\newcommand{\ee}{\end{equation}}

\newcommand{\tr}{\mathrm{Tr}}

\expandafter\let\csname equation*\endcsname\relax
\expandafter\let\csname endequation*\endcsname\relax

\begin{document}

\title{Entanglement Hamiltonian Tomography in Quantum Simulation}

\author{Christian Kokail}
\affiliation{Center for Quantum Physics, University of Innsbruck, Innsbruck, Austria}
\affiliation{Institute for Quantum Optics and Quantum Information of the Austrian Academy of Sciences, Innsbruck, Austria}

\author{Rick van Bijnen}
\affiliation{Center for Quantum Physics, University of Innsbruck, Innsbruck, Austria}
\affiliation{Institute for Quantum Optics and Quantum Information of the Austrian Academy of Sciences, Innsbruck, Austria}

\author{Andreas Elben}
\affiliation{Center for Quantum Physics, University of Innsbruck, Innsbruck, Austria}
\affiliation{Institute for Quantum Optics and Quantum Information of the Austrian Academy of Sciences, Innsbruck, Austria}
	
\author{Beno\^it Vermersch}

\affiliation{Center for Quantum Physics, University of Innsbruck, Innsbruck, Austria}
\affiliation{Institute for Quantum Optics and Quantum Information of the Austrian Academy of Sciences, Innsbruck, Austria}
\affiliation{Univ.  Grenoble Alpes, CNRS, LPMMC, 38000 Grenoble, France}	
	
\author{Peter Zoller}
\affiliation{Center for Quantum Physics, University of Innsbruck, Innsbruck, Austria}
\affiliation{Institute for Quantum Optics and Quantum Information of the Austrian Academy of Sciences, Innsbruck, Austria}

 \begin{abstract}
Entanglement is the crucial ingredient of quantum many-body physics, and characterizing and quantifying entanglement in closed system dynamics of quantum simulators is an outstanding challenge in today's era of intermediate scale quantum devices. Here we discuss an efficient tomographic protocol for reconstructing reduced density matrices and entanglement spectra for spin systems. The key step is a parametrization of the reduced density matrix in terms of an entanglement Hamiltonian involving only quasi local few-body terms. This ansatz is fitted to, and can be independently verified from, a small number of randomised measurements. The ansatz is suggested by Conformal Field Theory in quench dynamics, and via the Bisognano-Wichmann theorem for ground states. Not only does the protocol provide a testbed for these theories in quantum simulators, it is also applicable outside these regimes. We show the validity and efficiency of the protocol for a long-range Ising model in 1D using numerical simulations. Furthermore, by analyzing data from $10$ and $20$ ion quantum simulators [Brydges \textit{et al.}, Science, 2019], we demonstrate measurement of the evolution of the entanglement spectrum in quench dynamics.
\end{abstract}

\maketitle 
%\textbf{
%Entanglement is the crucial ingredient of quantum many-body physics, and characterizing and quantifying entanglement in closed system dynamics of quantum simulators is an outstanding challenge in today's era of intermediate scale quantum devices. Here we discuss an efficient tomographic protocol for reconstructing reduced density matrices and entanglement spectra for spin systems. The key step is a parametrization of the reduced density matrix in terms of an entanglement Hamiltonian involving only quasi local few-body terms. This ansatz is fitted to, and can be independently verified from, a small number of randomised measurements. The ansatz is suggested by Conformal Field Theory in quench dynamics, and via the Bisognano-Wichmann theorem for ground states. Not only does the protocol provide a testbed for these theories in quantum simulators, it is also applicable outside these regimes. We show the validity and efficiency of the protocol for a long-range Ising model in 1D using numerical simulations. Furthermore, by analyzing data from $10$ and $20$ ion quantum simulators [Brydges \textit{et al.}, Science, 2019], we demonstrate measurement of the evolution of the entanglement spectrum in quench dynamics.
%}

Quantum simulation realizes an isolated quantum many-body system with
dynamics governed by a designed many-body Hamiltonian ${H}$ \cite{georgescu2014,Browaeys2020,Monroe2019,Brydges2019,Kokail2019,wilkinson2020,King2018}. The aim
of quantum simulation is to study and characterize equilibrium states
and phases, and non-equilibrium dynamics of this artificial quantum
matter including their entanglement properties \cite{Cirac2012,Amico2008,zeng2019}. The Hamiltonian $H$
plays a unique role in determining these physical states of matter:
either as ground state, ${H}\ket{\Psi_{G}}=E_{G}\ket{\Psi_{G}}$, as
a finite temperature state in the form of a Gibbs ensemble $\sim\exp{(-\beta {H})}$;
or as generator of the quench dynamics with an initial (pure) state
$\ket{\Psi_{0}}$ evolving in time as $\ket{\Psi_{t}}=\exp{(-i{H}t)}\ket{\Psi_{0}}$.
Physical Hamiltonians, however, consist of a small set of terms with
quasi-local few-body interactions. Thus, for a given ${H}$, only a
small set of physical parameters determines the accessible quantum
states and their entanglement structure.

For an isolated quantum system in a pure state $\ket{\Psi}$, the
entanglement properties of a bipartition \mbox{$A\!:\!B$} are encoded in the Schmidt decomposition,
\mbox{$ 
\ket{\Psi}=\sum_{\alpha=1} ^{\chi_A} \lambda_\alpha \ket{\Phi^\alpha_{A} } \otimes \ket{\Phi^\alpha_{B} } 
$}  \cite{Amico2008}. 
Here, $\lambda_\alpha$ are Schmidt coefficients, and $\ket{\Phi^\alpha_{A} }$ are eigenvectors of the reduced density matrix,
\begin{equation}
{\rho}_{A}\equiv\exp\left(-\tilde{H}_{A}\right)=\sum_{\alpha=1}^{\chi_{A}}e^{-\xi_{\alpha}}\ket{\Phi_{A}^{\alpha}}\bra{\Phi_{A}^{\alpha}}\;,\label{eq:rhoA}
\end{equation}
with the Schmidt rank $\chi_A$ as proxy of entanglement. Eq.~(\ref{eq:rhoA}) defines
the entanglement (or modular) Hamiltonian (EH) $\tilde H_A$, and the  entanglement spectrum (ES) $\{\xi_\alpha\}$ via $\lambda_\alpha^2\equiv e^{-\xi_\alpha}$.
%\cite{peschel2009reduced}
The EH, and its spectrum and eigenvectors, thus fully specify
the entanglement properties of the many-body wave function for the bipartition \mbox{$A:B$}. They are crucial for our understanding of the role of entanglement in quantum many-body systems, with applications ranging from the detection and characterization of topological order and quantum phase transitions to  characterization of non-equilibrium phenomena and thermalization in closed quantum systems \cite{Regnault2015,Dalmonte2018,Zhu2020,Chang2019,Calabrese2016,Wen2018}. In addition, the ES and in particular $\chi_{A}$ determine the applicability of numerical techniques based on tensor networks \cite{orus2019}.

The EH $\tilde H_A$ can in principle be found via quantum state tomography (QST)
of $\rho_{A}$ \cite{Acharya2019,Gross2010,Cramer2010,Smolin12,Torlai2018, Choo2018}. However, without strong assumptions on the state~\cite{Cramer2010,Torlai2018}, the required number of measurements for QST  scales at least as $\chi_{A}2^{N_{A}}$, i.e.~exponentially in the subsystem
size $N_{A}$ and with Schmidt rank $\chi_{A}$  \cite{Haah2017,Wright2016, fern2020fast}.  Obtaining $\tilde H_A$ for highly entangled states $|\Psi\rangle$, i.e.~highly mixed ${\rho}_A$, generated in quantum simulation, is thus an outstanding challenge.  

\begin{figure*}[t]
	\centering
	\includegraphics[width=0.97\textwidth]{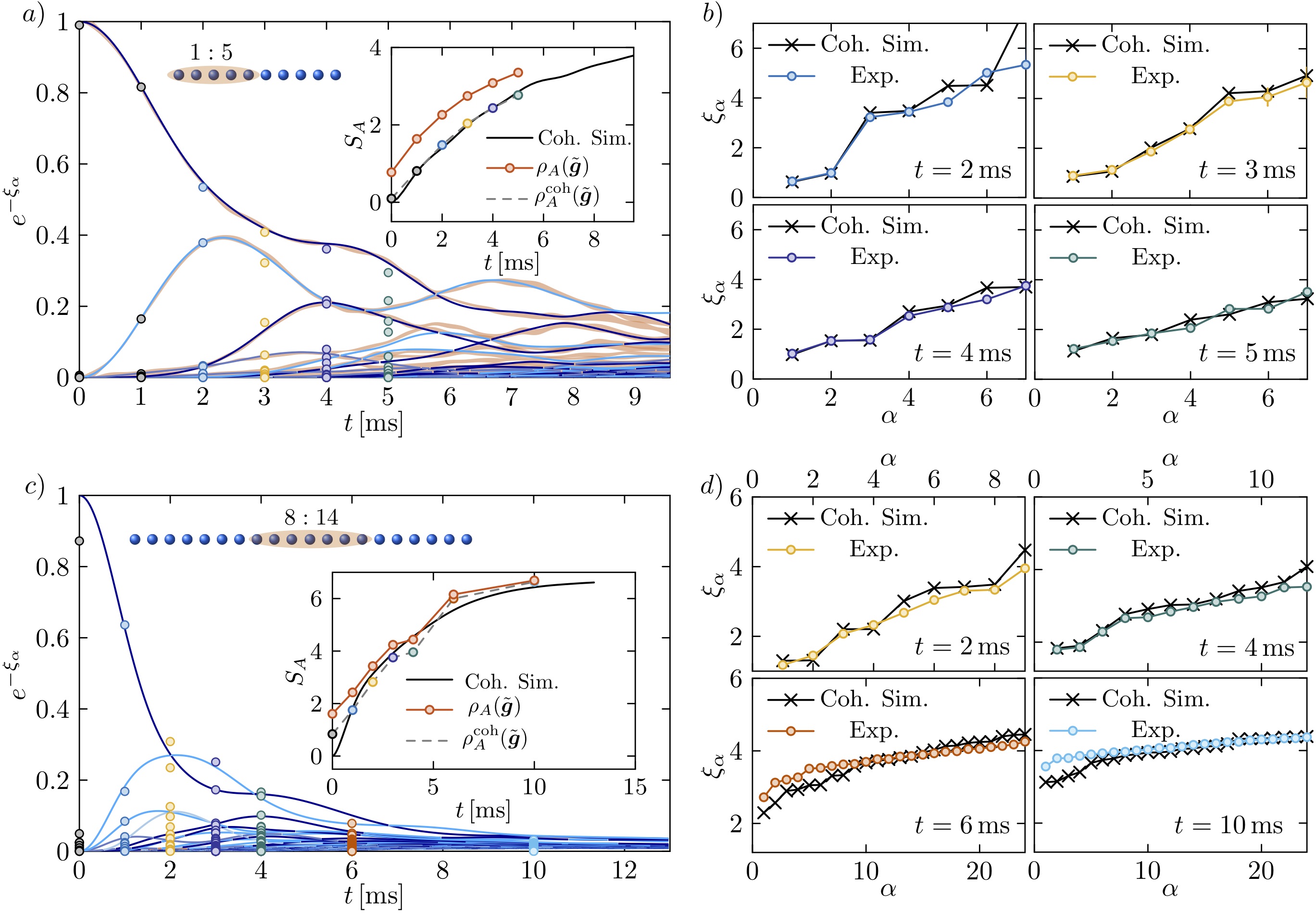}
	\caption{\textit{EHT in experimental quench dynamics}: Time evolution of the ES $\{\lambda_\alpha^2 \equiv e^{-\xi_\alpha}\}$ in Hamiltonian quench dynamics with a long-range transverse-field Ising model (\ref{eq:Ising}) ($B\gg J$) from an initial Nèel state $\ket{\uparrow \downarrow \uparrow \dots}$ on $10$ and $20$-spin trapped-ion quantum simulators, for $5$ and $7$-spin partitions, as shown in the insets. a), c) ES as a function of time.  Colored circles show the eigenvalues of $\rho_A^{\text{coh}}(\tilde{\boldsymbol{g}})$ (defined in Eq.~(\ref{eq:depol}) in the), obtained from EHT on experimental data with $N_M = 150$ measurements taken in $N_U=500$ different random bases. Solid lines show theoretical simulations of the Hamiltonian quench dynamics, with different colors indicating the total magnetization of the Schmidt components. Orange colored bands in a) show the $68\%$ confidence interval obtained by repeatedly simulating the full EHT procedure with the same settings on the entire time interval. Insets in a) and c): von Neumann entanglement entropy \mbox{$S_A = - {\rm Tr} (\rho_A \log_2 \rho_A )$} as a function of time. Black solid line indicates exact results from simulation of coherent Hamiltonian dynamics. Colored points show $S_A$ extracted from EHT on experimental data, obtained from the spectrum of $\rho_A^{\text{coh}}(\tilde{\boldsymbol{g}})$, while orange points show $S_A$ obtained from  the full ansatz (see Eq.~(\ref{eq:depol}) in Appendix  \ref{methods:EHT}) including imperfect initial state preparation and measurement errors. b), d) Reconstructed lowest eigenvalues $\{\xi_\alpha\}$ of the EH in EHT at later times, and comparison with ES obtained from numerical simulations of exact Hamiltonian dynamics. Error bars, calculated via Jackknife resampling, are mostly smaller than symbols.}
	\label{fig:Exp}
\end{figure*}

Our goal is the development of experimentally feasible measurement protocols to \emph{directly} and \emph{efficiently} determine ${\tilde{H}_{A}}$ for quantum states generated in  quantum simulation.  We approach the problem by choosing a physically motivated ansatz for $\tilde H_A$ consisting of a small set of quasi-local few-body terms with variable parameters $\tilde{\boldsymbol{g}}$. The parameters are fitted to experimental data, approximating ${\rho}_{A}$ with $\rho_A(\tilde{\boldsymbol{g}})\sim \exp [-\tilde H_A(\tilde{\boldsymbol{g}})]$. We note that $\rho_A(\tilde{\boldsymbol{g}})$ has the form of a Gibbs ensemble, for which a formal proof for efficient sampling can be found \cite{anshu2020sampleefficient}. In Appendix \ref{methods:EHT}, we describe the entire Entanglement Hamiltonian Tomography (EHT) protocol in more detail. In brief, the protocol applies $N_U$ independent single-qubit rotations, ${U}=\bigotimes_{i\in A}{u}_{i}$ and ${u}_{i}$  sampled from a unitary 2-design \cite{PhysRevA.80.012304,Huang20}, to the quantum state, followed by a read out of spin states, repeated $N_M$ times. 

The complexity of the ansatz $\tilde H_A(\tilde{\boldsymbol{g}})$ depends on the physical system, and can be systematically expanded and verified at each step of the process, possibly unveiling new physics in the underlying quantum state. At the base level, the ansatz consists of the system Hamiltonian restricted to the subsystem $A$, but with spatially varying coefficients: for ${H}=\sum_{\{i\}}h_{\{i\}}$ being a sum of quasi-local terms $h_{\{i\}}$ acting on some
neighborhood of lattice sites $\{i\}$, we write for the entanglement Hamiltonian ansatz $\tilde{H}_{A}(\tilde{\boldsymbol{g}})=\sum_{\{i\}\in A}\tilde{g}_{\{i\}}\,h_{\{i\}}+{\mathcal{K}}_{A}$ with parameters $\tilde{g}_{\{i\}}$. Possible extensions and corrections $\mathcal{K}_A$, with corresponding parameters, can be added when required (to be discussed below). Importantly, EHT allows for directly verifying the correctness of the ansatz, by computing cross-correlations between the experimental observations and predictions from $\rho_A(\tilde{\boldsymbol{g}})$ (see Appendix \ref{Methods:XP}, and \cite{ElbenXPlatform}). 

A physical motivation for the quasi-local ansatz involving few-body terms for EH comes from the Bisognano--Wichmann (BW) theorem of axiomatic quantum field theory \cite{BW1,BW2}, and from Conformal Field Theory (CFT) \cite{Calabrese2009}, and is implicit in the Li - Haldane conjecture \cite{PhysRevB.86.045117, Regnault2015}. The BW theorem provides a closed form expression for EH for ground states in systems with Lorentz invariance, valid in all spatial dimensions \cite{BW1,BW2}. This theorem states that, given a system with a Hamiltonian density ${\cal H}(\mathbf{x})$ and a half-partition $A$ of the infinite plane (denoted here for simplicity as $x>0$), the EH of the ground state reads $\tilde{H}_{A}=2\pi\int_{x\in A}d\mathbf{x}\left[x{\cal H}(\mathbf{x})\right]+c'$, where $c'$ is a normalization constant. The BW theorem thus predicts that the EH of the ground state is built from just local and few body terms, as appearing in the original Hamiltonian, and the reduced density matrix to have the structure of a Gibbs state ${\rho}_{A}\sim\exp\left[-\int_{x\in A}d\mathbf{x} \, \beta(x){\cal H}(\mathbf{x})\right]$ with a spatially dependent local inverse `temperature' $\beta(x)\sim x$.
While the BW theorem applies strictly speaking only to infinite systems and the continuum, although in all spatial dimensions, BW is readily adapted to finite size lattice models. Numerical simulations demonstrate remarkable ability to predict the ES for a range of interacting many-body models~\cite{Dalmonte2018} (see also \cite{Alba2012, Eisler_2017, PhysRevLett.58.1395, PhysRevB.98.134403}). 

In a similar way, CFT makes specific predictions in $1+1$d about the structure of the EH for ground states, and for quantum quenches to a critical point \cite{hislop1982, Cardy2016,Wen2018}. 
 Consistent with a quasi particle picture, this EH has again a local structure with contributions from the energy (Hamiltonian) and a momentum density (see  Appendix \ref{supp:CFT}). This suggests corrections to the above deformed Hamiltonian ansatz for the EH, i.e.~physical predictions which can be tested with EHT. For long times, CFT describes the emergence of a local thermal equilibrium in form of a (generalized) Gibbs ensemble.  EHT thus provides a direct and fine-grained testbed for  BW and CFT predictions in quantum simulation experiments. We emphasize, however, that a quasilocal EHT ansatz can also be applied, tested and verified in regimes outside the immediate validity of BW and CFT, as discussed below, as well as for higher spatial dimensions and in the presence of local noise and decoherence  (see Supplementary Information). 

In Fig.~\ref{fig:Exp} we demonstrate the EHT protocol in an experimental setting \cite{Brydges2019}, extracting the time evolution of the Schmidt spectrum 
$\{\lambda_\alpha^2 \equiv e^{-\xi_\alpha}\}$ 
in quench dynamics on $10$ and $20$-spin trapped-ion quantum simulators for $5$ and $7$-qubit partitions. The analog quantum simulator implements a long-range transverse Ising model \cite{Monroe2019,Brydges2019}, where an initial product state evolves towards a highly entangled state. Figs.~\ref{fig:Exp} a) and c) demonstrate the ability of EHT to faithfully extract the Schmidt spectrum, and thus the Von Neumann entropy, from a small number of measurements, while Figs.~\ref{fig:Exp} b) and d) show excellent agreement of the experimentally inferred ES with the theoretical predictions. Moreover, employing the verification protocol described in \cite{ElbenXPlatform}, we are able to experimentally verify the fidelity of our theoretical reconstruction of $\rho_{A}$ with an independently taken experimental data set (see Appendix \ref{Methods:XP}). Our analysis  is based on existing experimental data sets published in \cite{Brydges2019} involving quench dynamics, followed by local random rotations and projective single spin measurements.

In the remainder of the paper we will discuss the construction of the ansatz and its operator content. We will develop and illustrate EHT in context of the 1D antiferromagnetic transverse
Ising model with Hamiltonian
\begin{equation}
 H_{I}=\sum_{i<j}J_{ij}\sigma_{i}^{x}\sigma_{j}^{x}+B\sum_{i}\sigma_{i}^{z}. \label{eq:Ising}
\end{equation}
Such a spin model is realized with trapped ions, as a long range Ising
model $J_{ij}= J /|i-j|^{\eta}$ with $0 \leq \eta \leq 3$  \cite{Monroe2019,Brydges2019,Kokail2019},
and for laser excited Rydberg atoms with Van der Waals interaction
$\eta=6$ \cite{Browaeys2020}. Our discussion below will proceed in three steps. We will first demonstrate EHT applied to Ising ground states by means of theoretical simulations, finding that we can compare directly with BW predictions with a moderate measurement budget. Second, we apply EHT to quenches to a critical point. 
We make an ansatz supported by CFT predictions to determine the EH and ES, and compare EHT runs to the exact theoretical results. This establishes EHT as framework in quantum simulation to observe and test features of BW and CFT in context of lattice models.
Finally, we return to EHT for quench dynamics in experimental trapped-ion quantum simulation already described in Fig.~\ref{fig:Exp}.

\begin{figure}[t]
	\centering
	\includegraphics[width=0.97\linewidth]{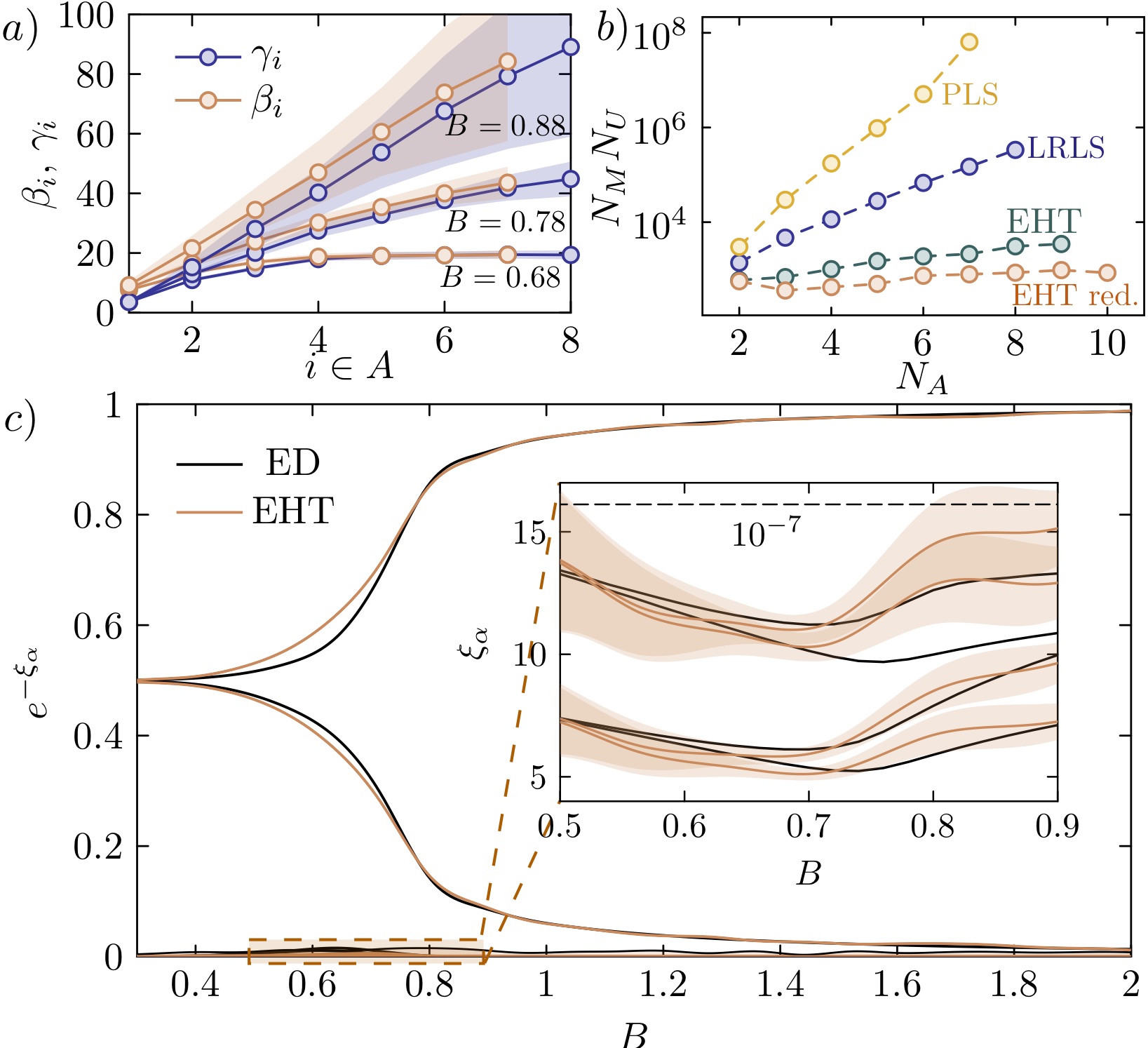}
      	\caption{\textit{Simulation of EHT for ground states of a long-range transverse field Ising  model.} a) Parameters $\beta_i$ and $\gamma_i$ obtained from fitting an EH ansatz with local independent paramters for 3 different values of the transverse field across the phase transition. The behaviour of the fit parameters reveals the existence of a local temperature and confirms the spatial dependence predicted by the BW theorem in the short range correlated phase. b) Required number of measurements in order to reach an Uhlmann fidelity of 99\% with respect to the exact density matrix, for EHT with $N_U = 1$ and local variational parameters (EHT), EHT with $N_U = 1$ and 3 parameters (EHT red.) (see main text) and 2-rank least squares (LRLS) with $N_U = 64$ c) Eigenvalues of the reduced density matrix for a subsystem of $N_A = 8$ sites on the right boundary of a 50-site long-range Ising chain obtained from EHT with $N_U = N_M = 150$ and from an exact Schmidt decomposition of the ground state (ED). The inset zooms in on the low-lying part of the Schmidt spectrum, showing Schmidt components up to a level of $10^{-7}$. Error bands are obtained from repeating the fitting 50 times with different random unitaries and computing the standard error.}
	\label{fig:TomoBW}
\end{figure}

\textbf{EHT for ground states \& BW lattice ansatz --} 
As a first demonstration of EHT we perform numerical experiments on ground states of a long-range Ising chain (\ref{eq:Ising}) with open boundary conditions and $\eta = 2.5$. For a subsystem $A$ of this system, we employ the base level ansatz with a deformed system Hamiltonian, i.e. \mbox{$\tilde{H}_A = \sum_{j<i \in A} \tilde{J}_{ij} \sigma_i^x \sigma_j^x + \sum_{i \in A} \tilde{B}_i \sigma_i^z$}, and $\mathcal{K}_A = 0$. We study the tomographically constructed EH as a function of the transverse field $B$,  crossing the phase transition between a paramagnetic phase and symmetry-broken phase with a closing entanglement gap. 
Here, EHT allows for an ab-initio test of  BW predictions for lattice models, and enables an efficient reconstruction of the reduced density matrix from very few measurements. 

We apply EHT to a subsystem of $N_A = 8$ sites at the edge of a $N=50$ system, parametrizing $\tilde{H}_A$ with independent local variational parameters for the interaction part $\tilde{J}_{ij} = J_{ij} \beta_i$ and the local part $\tilde{B}_i = B \gamma_i$. We simulate the procedure by fitting the EH parameters to $N_M = 5\cdot10^3$ samples taken from the ground state, in each of $N_U = 20$ random measurement bases, and for various values of the local field $B$ around the phase transition. Fig.~\ref{fig:TomoBW} shows that the resulting EH parameters $\boldsymbol{\beta}$ and $\boldsymbol{\gamma}$ coincide within the error bars, $\boldsymbol{\beta}=\boldsymbol{\gamma}$, revealing the existence of a local temperature, which shows that EHT in an experiment provides a framework to test BW predictions. In the paramagnetic phase, $B = 0.88J$, the system exhibits exponentially decaying correlations with a correlation length much smaller than the subsystem size. Consequently, the magnitude of the fitted EH parameters increases linearly with the distance to the entanglement cut, in accordance with  BW expectations.
In the antiferromagnetic phase, the correlation lengths far exceed the subsystem size, and we find that the EH parameters deviate from BW. In this region, EHT reveals an additional quadratic spatial variation.

\begin{figure}[t]
	\centering
	\includegraphics[width=0.99\linewidth]{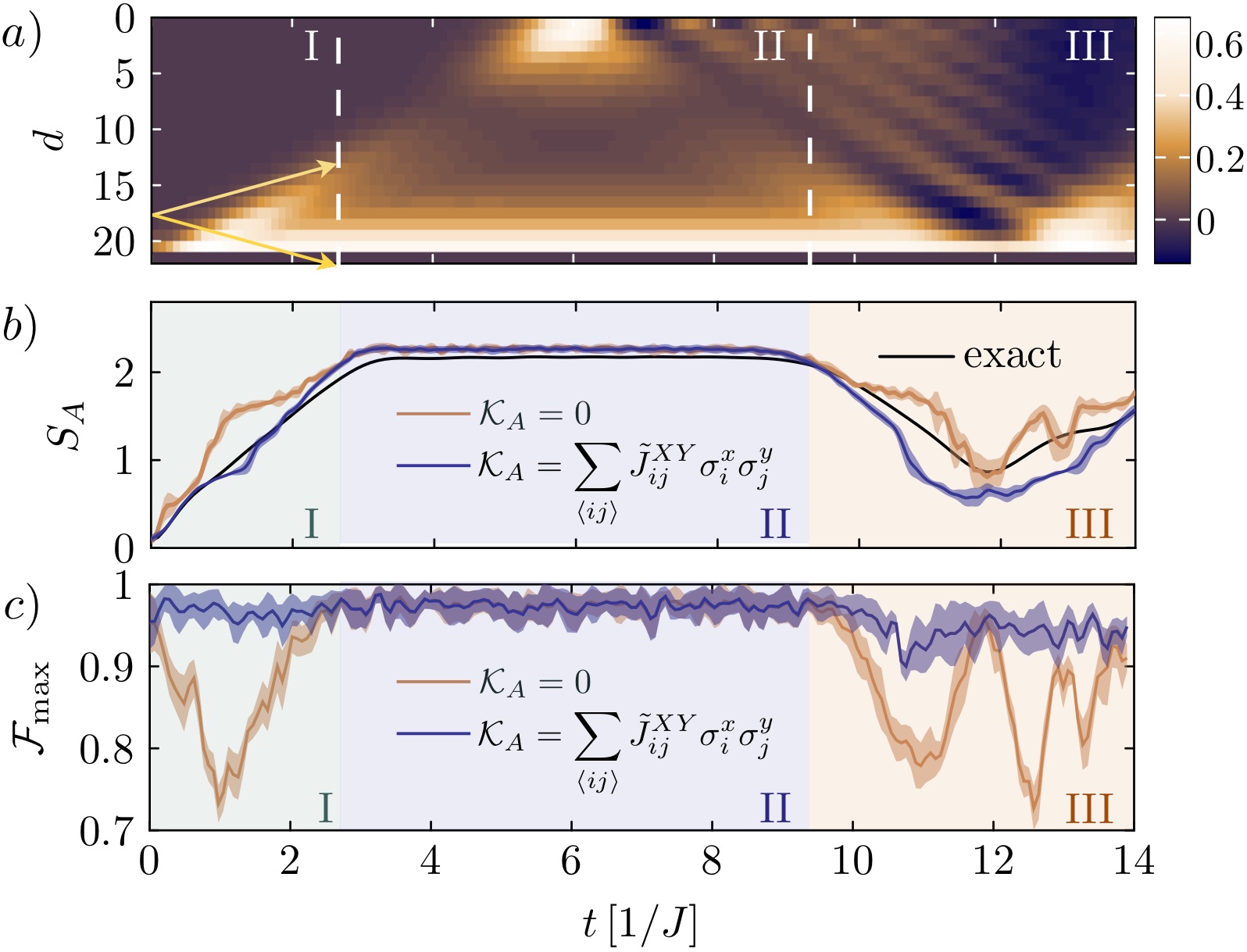}
	\caption{\textit{Simulation of EHT for a global quench in the critical Ising model.} a) Spin-spin correlation functions $\braket{\sigma^z_1 \sigma_d^z} - \braket{\sigma^z_1}\braket{\sigma^z_d}$ as a function of time. Consistent with a quasi-particle picture, two spins at site $1$ and $d$ become correlated at time $c t=d/2$ due to a pair of quasi-particles generated at site $d/2$ at $t=0$ which travel in opposite directions with speed $c/J \approx \pm 2$ (yellow arrows). b) von Neumann entropies for $N_A = 5$ as a function of time obtained from EHT with and without momentum terms included in the ansatz and with $N_U=N_M=150$. Error bands are obtained by calculating the standard deviation after repeating the fitting 100 times. (c) Discovery of momentum contributions by measuring the maximum fidelity $\mathcal{F}_\text{max}(\rho_A, e^{-\tilde{H}_A(\tilde{\boldsymbol{g}})})$ between the exact density matrix $\rho_A$ and $e^{-\tilde{H}_A(\tilde{\boldsymbol{g}})}$ obtained from the EHT. The error band is estimated with $N_U=500$ and $N_M=150$. Including momentum terms leads to a significant fidelity enhancement in regions I and III.}
	\label{fig:CFT}  
\end{figure} 

The latter observation suggests an EH ansatz with fewer variational parameters, where we choose a parabolic deformation of the EH coefficients $\tilde{J}_{ij} = J_{ij}  \sum_{k=0}^2 \beta_k (i+j-1)^k$ and $\tilde{B}_i = B \sum_{k=0}^2 \beta_k (i-1/2)^k$.
This ansatz depends only on $3$ variational parameters, independent of the subsystem size. In Fig.~\ref{fig:TomoBW} we simulate the EHT procedure while scanning $B$ across the phase transition.  The extracted entanglement spectrum clearly exhibits the phase transition in the entanglement gap, and reproduces the smaller eigenvalues down to $\sim 10^{-5} - 10^{-7}$ accuracy. 

The low number of fitting parameters allows for an efficient tomographic procedure requiring very few measurements. Fig.~\ref{fig:TomoBW} b) shows the scaling for the total number of measurements $N_U \times N_M$ in order to achieve an Uhlmann fidelity $\mathcal{F}(\rho) = \left[ \text{Tr} (\sqrt{\sqrt{\rho_A} \rho \sqrt{\rho_A}}) \right]^2$ exceeding 99\% with respect to the exact (theoretical) density matrix, as a function of the size of the subsystem $N_A$, and for all values of the transverse field. A similar analysis for highly mixed Gibbs states is shown in the Supplementary Information.
%The reconstructed density matrices achieve a fidelity exceeding $99.6\%$ everywhere. 
For comparison, the scaling is shown for two competing tomographic methods, projected least squares (PLS)  \cite{Guta20, Huang20} and low rank least squares (LRLS) \cite{Riofrio2017} (see Appendix \ref{Methods:QST}). Remarkably, EHT for this particular example requires measurements only in a single basis. Typically, tomographic protocols only provide estimates for the diagonal elements of the density matrix in a given measurement basis, thus requiring multiple measurement bases in order to determine off-diagonal elements. In contrast, in EHT the relation between diagonal and off-diagonal elements is fixed by the ansatz, and measurements of the diagonal suffice to determine the entire matrix.

\textbf{EHT for quench dynamics near criticality --} Here we test and verify a quasi-local ansatz for the EH in quench dynamics by simulating EHT measurement runs. We focus in particular on global quenches in critical lattice models, where such an ansatz is suggested by CFT.  Here, EHT allows not only for the reconstruction of such states near criticality with very few measurements, but terms beyond the base level  EH ansatz ($\mathcal{K}_A \neq 0$) also reveal underlying physical phenomena connected to entanglement growth and quasi-particle spreading (see also \cite{Elben2020negativity}). 

We consider the Ising model (\ref{eq:Ising}) with nearest-neighbour interactions, i.e.\ $\eta \rightarrow \infty$. Starting with a short-range entangled ground state in the gapped paramagnetic phase at $B=2.5 J$, we perform a quantum  quench of the transverse field to a value $B_c = 0.97 J$ close to the critical point, predominantly populating the low energy part of the many-body spectrum.  The ansatz for the time-dependent EH  is motivated by the  (continuum) EH obtained in CFT which provides an effective low-energy description of the critical Ising model \cite{Calabrese2016}. As elaborated in  the Supplementary Information,  the  EH in a CFT is composed of  energy $T_{00}(x) = \mathcal{H}(x)$ and momentum $T_{01}(x)= \mathcal{P}(x)$ densities.
Their lattice analogs can be defined as \mbox{$ h_i = J_{i,i+1} \sigma_i^x \sigma_{i+1}^x + B \sigma_i^z$} and \mbox{$p_i = i [ h_{i+1}, h_{i} ] \sim \sigma_{i+1}^x \sigma_{i}^y$}, respectively \cite{PhysRevB.96.245105}.
Thus, we choose  EH ansätze of the form \mbox{$\tilde{H}_A = \sum_{i \in A} \left( \tilde{J}_{i, i+1} \sigma_i^x \sigma_{i+1}^x + \tilde{B}_i \sigma_i^z \right) + \mathcal{K}_A$}, with free fit parameters  $\tilde{J}_{i, i+1}$ and  $\tilde{B}_{i}$. Here, $\mathcal{K}_A$ accounts for possible momentum contributions.

 In Fig.~\ref{fig:CFT}, we compare the performance of the base level EHT ansatz (i.e.\ $\mathcal{K}_A\equiv 0$) and with an EHT ansatz where momentum contributions \mbox{$\mathcal{K}_A = \sum_{\braket{i,j} \in A} \tilde{J}_{i, j}^{XY} \sigma_i^x \sigma_{j}^y$} with free fit parameters $\tilde{J}_{i, j}^{XY}$  have been added. Here, $\braket{i,j}$ denotes nearest-neighbor sites. We choose a subsystems of  $N_A=5 $ qubits at the edge of a total system with $N=22$ lattice sites. We sample  $N_M = 150$ projective measurements in $N_U=150$ measurement settings  from the simulated quantum state, to which both ansätze are fitted according to Eq.~(\ref{eq:method}). 
In accordance with a picture of long-lived pairs of quasi-particles \cite{Calabrese2016}, generated at $t=0$ and traveling with speed $c=2J$ [see Fig.~\ref{fig:CFT} a)], we observe three distinct regions in the time evolution. 

\begin{figure}[t]
	\centering
	\includegraphics[width=0.99\linewidth]{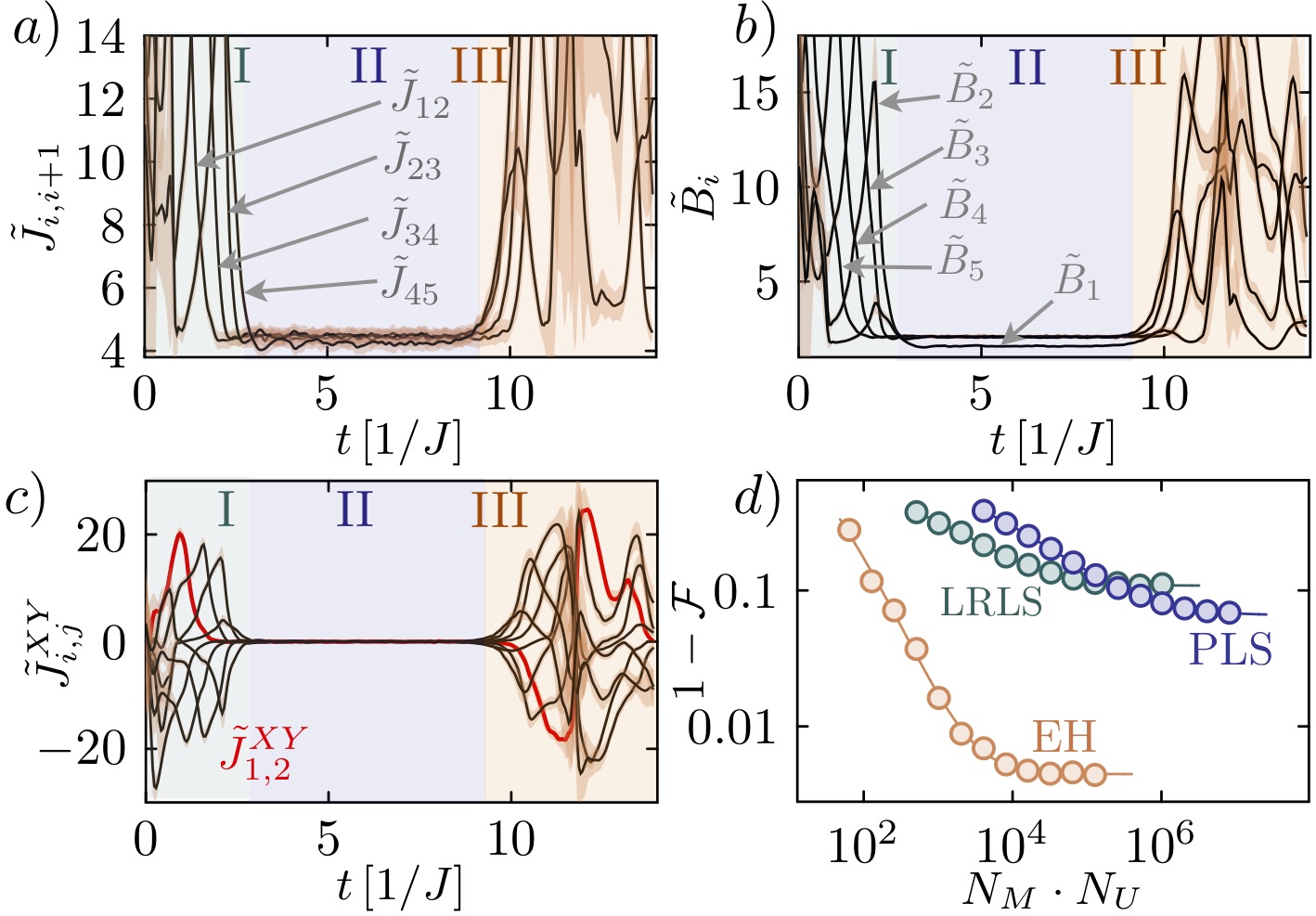}
	\caption{\textit{EH local parameters obtained from simulation of EHT, and fidelity for quench dynamics} in a critical Ising chain a)-c) Parameters of the fitted EH, where $\tilde{J}_{i, i+1}$ and $\tilde{B}_{i}$ correspond to terms present in the energy, and $\tilde{J}_{i, j}^{XY}$ account for lattice momenta. Panel c) highlights the coefficient $J_{1,2}^{XY}$ showing the different signs of the couplings depending on the propagation direction of quasi particles.  Panel d) shows EHT  reaching a significantly smaller  infidelity of the estimated density matrix $\rho_A(\tilde{\boldsymbol{g}})$ at time $Jt=7$ with less simulated experimental runs $N_M N_U$ than other tomographic techniques.}
	\label{fig:CFTCouplings}  
\end{figure} 

At early times, in region I,  the entanglement entropy [Fig.~\ref{fig:CFT} panel b)] is increasing linearly. The inclusion of momentum terms in the EHT ansatz leads to a significantly improved fidelity of the estimated density matrix [Fig.~\ref{fig:CFT} panel c)]. The couplings $\tilde{J}_{i, i+1}$,  $\tilde{J}_{i, j}^{XY}$ and  fields $\tilde{B}_{i}$  are of similar magnitude and dynamically oscillating [Fig.~\ref{fig:CFTCouplings} (a-c)].  In addition, we show in the Supplementary Information that momentum contributions of the form $\sigma_i^x \sigma_j^y$ are the only type of 2-body terms that lead to a significant fidelity enhancement in region I, when added to the EH ansatz. This is consistent with CFT predicitions for non-vanishing quasi-particles currents entering subsystem $A$ \cite{Calabrese2016,Zhu2020}, and thus demonstrates   the ability of EHT for the discovery of physical phenomena.
 
At $ct\approx N_A$, the entanglement entropy saturates, and remains constant throughout region II. Here, both ansätze yield the same fidelity  [Fig.~\ref{fig:CFT} panel c)],
%the momentum contributions vanish  
and the entanglement Hamiltonian is determined by the energy densities $h_i$ only, multiplied with an effective, inverse temperature $\beta_i$ [Fig.~\ref{fig:CFTCouplings} panels (a-c)].  Consistent with the CFT prediction \cite{Cardy2016,Wen2018,Zhu2020} and the eigenstate thermalization hypothesis \cite{Deutsch1991,Srednicki1994,Rigol2008,Garrison2018}, $\beta_i$ is constant far from the entanglement cut, and decreasing towards the boundary. 

In region III, we observe a  dip in the entanglement entropy at $ct\approx N$ which   reflects the (approximate) revival of the initial state due to the  finite system size \cite{Calabrese2016, Modak_2020}.  At this time, pairs of quasi-particles that have been reflected at the boundaries meet again at their origins. The EH parameters show  dynamical oscillations and non-vanishing momentum terms accounting for traveling quasi-particles leaving the subsystem $A$. 
In this region, both  fidelities shown in Fig.~\ref{fig:CFT} are decreasing, whereby the ansatz including momentum terms still performs significantly better.
This shows that even an ansatz including both energy  and momentum densities can not fully represent the actual entanglement Hamiltonian in region III. In principle, the fidelity of the reconstructed density matrix can be enlarged via a systematic inclusion of additional terms, thereby discovering the relevant higher order terms in the EH.
This highlights the potential of EHT to provide insights on form and structure of the EH, in regimes without theoretical predictions.

\textbf{EHT in experimental quench dynamics --}  Finally, we return to the quench experiments already described in Fig.~\ref{fig:Exp}, a system for which CFT is \textit{a priori} not applicable. Nevertheless, we will demonstrate that EHT provides a systematic way of constructing a verifiable estimate of the reduced density matrix, with an EH built from quasi-local few-body operators. The experimental data  was taken in a trapped-ion quantum simulation experiment for quench dynamics with $N=10$ to $20$ spins~\cite{Brydges2019}. In this experiment, an initial product state was prepared as a Néel state $\ket{\uparrow\downarrow\uparrow\dots}$, with subsequent time evolution under the transverse field Ising Hamiltonian (\ref{eq:Ising}) with $\eta\approx1.24$ and $B\gg J$, effectively implementing magnetisation conserving exchange interactions, \mbox{$H_{I}=\sum_{i<j}J_{ij}\sigma_{i}^{+}\sigma_{j}^{-}+{\rm H.c.}+B\sum_{i}\sigma_{i}^{z}$}. A large amount of data was collected in randomized measurement bases at various points in time, in order to measure second order Renyi entropies. We now apply EHT to these datasets, while we refer to Appendix \ref{app:decoh} for technical details of how to adapt EHT to account for decoherence and measurement imperfections.

\begin{figure}[t]
	\centering
	\includegraphics[width=0.99\linewidth]{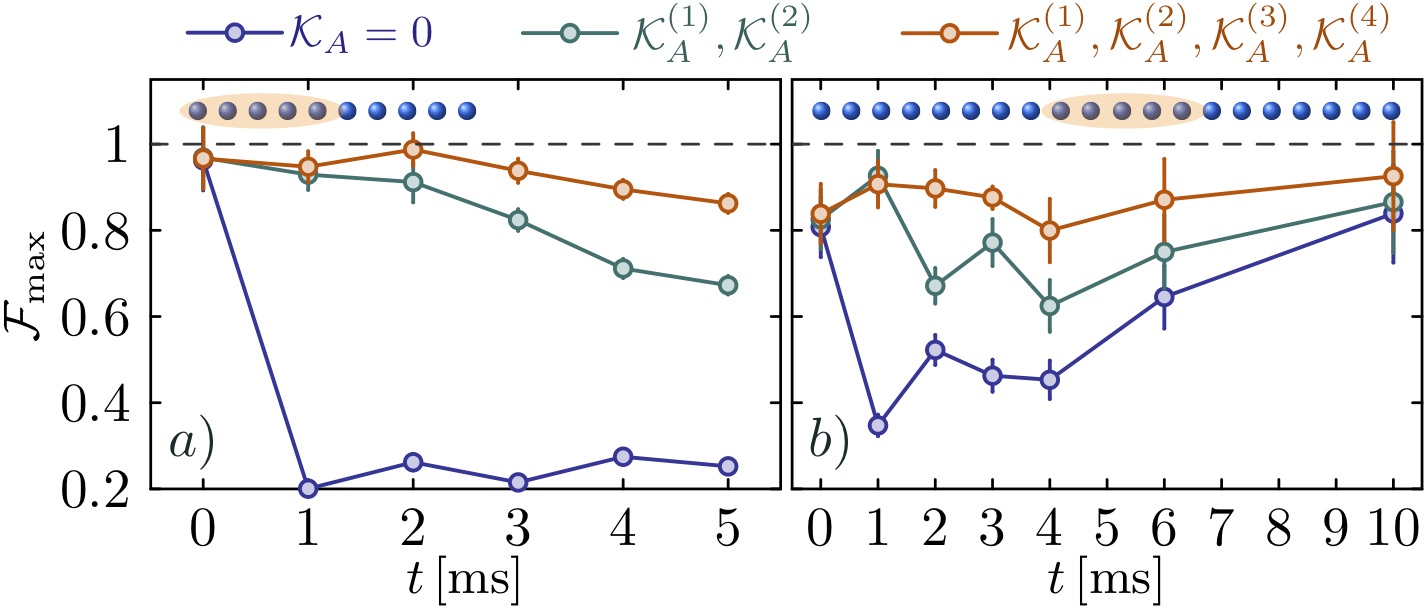}
	\caption{\textit{Experimental verification of EHT in quench dynamics on 10 and 20-spin trapped-ion quantum simulators --} Maximum fidelity ${\cal F}_{\rm max}$ (see Appendix \ref{Methods:XP}) of the tomographically reconstructed density matrix $\rho_A(\tilde{\boldsymbol{g}})$ (see Eq.~(\ref{eq:depol}) in Appendix \ref{app:decoh}, with $p=0.039$ (10 ions) and $p=0.070$ (20 ions) inferred from the initial state) for subsystems of 5 spins (shown as insets) with respect to a distinct data set for different ansatz complexities, starting from a deformed system Hamiltonian $H_A$, and subsequently including correction terms $\mathcal{K}_A$ listed in Appendix \ref{Methods:K}. EHT was performed on a data set consisting of $N_U = 200$ unitaries and the maximum fidelity was estimated with respect to the remaining data (300 unitaries). Error bars are obtained by Jackknife resampling.}
	\label{fig:Verification}  
\end{figure} 

In Fig.~\ref{fig:Verification} we consider EHT on $10$ and $20$ ion data with ans\"atze of the form $\tilde{H}_A(\tilde{\boldsymbol{g}}) = \sum_{ij \in A} \left( \tilde{J}_{ij}  \sigma_i^+ \sigma_j^- + \text{H.c.} \right) + \sum_{i \in A} \tilde{B}_i \sigma_i^z + \mathcal{K}_A$. We test various levels of ansatz complexity on the data, each time evaluating its performance with the verification protocol described in Appendix \ref{Methods:XP}. The straightforward deformation of the system Hamiltonian (i.e., $\mathcal{K}_A = 0$) yields rather low fidelities, suggesting that additional terms need to be added. We consider momentum terms  (see $\mathcal{K}^{(1)}_A, \mathcal{K}^{(2)}_A$ in Eqs. (\ref{EqKA1}, \ref{EqKA2}) in Appendix \ref{Methods:K}), as suggested by CFT (see Fig.~\ref{fig:CFTCouplings}),  improving the fidelity to above $70\%$ almost everywhere. Further improvement can be gained by adding corrections $\mathcal{K}^{(3)}_A, \mathcal{K}^{(4)}_A$ consisting of all possible $2$ and $3$ body operators that conserve the magnetization in the subsystem (see Eqs. (\ref{EqKA3}, \ref{EqKA4}) in Appendix \ref{Methods:K}), as we expect the reduced density matrix to be block-diagonal in the various magnetization subsectors, on behalf of the global Hamiltonian conserving that symmetry. The latter two additions boost the experimentally verified fidelity into the $80\% - 95\%$ regime for the 20-ion dataset, and fidelities in the $94\% - 99\%$ regime for the 10-ion data in Fig.~\ref{fig:Exp} and Fig.~\ref{fig:Verification}. These high fidelities allowed also for the determination of the von Neumann entropy, as shown in Fig.~\ref{fig:Exp}, in excellent agreement with theory simulations. 

\textbf{Conclusions and Outlook --} EHT based on a quasi-local ansatz for the EH provides an efficient technique to determine EH from few measurements for ground states and quench dynamics of lattice models, on behalf of a polynomial number of fitting parameters.  While such a parametrization is originally suggested by physics underlying BW for ground states and CFT for quenches to the critical point, this ansatz can be systematically extended, tested and experimentally verified -- or falsified -- in a much broader setting. This includes generalizations to higher spatial dimensions, and bosonic and fermionic systems, whereas we do not expect to cover states without efficient few-parameter description, as generated for instance by random quantum circuits \cite{arute2019quantum}. We have demonstrated the application of the protocol to experimental data for quench dynamics from a trapped-ion simulator, allowing us to extract entanglement properties including high-precision ES and Von Neumann entanglement entropies. The present results provide opportunities for a systematic and scalable experimental study of entanglement properties for equilibrium phases  and non-equilibrium phenomena in today's quantum simulators \cite{Browaeys2020,Monroe2019,wilkinson2020}.

\textit{Acknowledgment -- } We thank P.~Calabrese, M.~Dalmonte, G.~Giudici, Lata K.~Joshi, B.~Kraus, R.~Kueng, C.~Roos, L.~Sieberer, Jinlong Yu, and Wei Zhu for discussions, and members of the Innsbruck trapped-ion group for generously sharing the experimental data of Ref.~\cite{Brydges2019}.  Work at Innsbruck is supported by the European Union program Horizon 2020 under Grants Agreement No.~817482 (PASQuanS) and No.~731473 (QuantERA via QTFLAG), the US Air Force Office of Scientific Research (AFOSR) via IOE Grant No.~FA9550-19-1-7044 LASCEM, by the Simons Collaboration on Ultra-Quantum Matter, which is a grant from the Simons Foundation (651440, PZ), and by the Institut f\"ur Quanteninformation. BV acknowledges funding from the Austrian Science Fundation (FWF, P 32597 N). The computational results presented here have been achieved (in part) using the LEO HPC infrastructure of the University of Innsbruck. Numerical calculations were performed (in part) using the ITensor library \cite{itensor}.

\appendix
\counterwithin{figure}{section}

\section{Entanglement Hamiltonian Tomography} \label{methods:EHT}
In EHT we construct an estimator of the state $\rho_A$ of the form $\rho_{A}(\tilde{\boldsymbol{g}})=\exp[-\tilde{H}_{A}(\tilde{\boldsymbol{g}})] / Z(\tilde{\boldsymbol{g}})$, with $Z(\tilde{\boldsymbol{g}}) = \tr\left(\exp[-\tilde{H}_{A}(\tilde{\boldsymbol{g}})]\right)$ a constant to ensure unit trace. The ansatz for the entanglement Hamiltonian $\tilde{H}_{A}(\tilde{\boldsymbol{g}})$ is constructed from quasi-local few-body operators, as detailed in the main text. The (polynomially many) real-valued coefficients $\tilde{\boldsymbol{g}}$ of each of the operators are inferred from quantum measurements on $\rho_A$.
%by matching predictions of the estimator in a least squares sense with experimental observations from \textit{local} measurements performed on independent single copies of the density matrix.
In particular, we consider measurements of bitstrings $s$ in the computational basis after randomized local rotations ${U}=\bigotimes_{i\in A}{u}_{i}$, where ${u}_{i}$ is a local basis rotation at site $i$, and is sampled from a unitary 2-design \cite{PhysRevA.80.012304}. Denoting by $P_U(\mathbf{s})$ the frequency of having observed a particular bitstring $\mathbf{s}$ in the experiment, one can define a cost function
\begin{align}
\chi^2 = \sum_U \sum_\mathbf{s} \left[ P_U(\mathbf{s}) - \tr\left( \rho_A{(\boldsymbol{\tilde g}}) U\ket{\mathbf{s}}\bra{\mathbf{s}}U^\dagger \right) \right] ^2, \label{eq:method}
\end{align}
which is to be minimized over the parameters $\tilde{\boldsymbol{g}}$ of the estimator.  Our protocol thus learns the quasi-local EH from data, in contrast to the learning of quasi-local system Hamiltonians  from pure (eigen-) states \cite{Qi2019,Bairey2019,Wang2017}. We note that  choices of cost-functions and  random unitary ensembles \cite{elben2018renyi} other than in Eq.~\eqref{eq:method} are possible, and devote their detailed investigation to future work.

Extracting the EH and ES from experimental data, as described in context of Fig.~(\ref{fig:Exp}), requires the EHT protocol outlined above to be adapted to account for decoherence and imperfections. First, the experimental time evolution will be weakly coupled to an environment, which in the present case is well modeled by a dephasing master equation \cite{Brydges2019,ElbenXPlatform}.
Second, experimental randomized measurements suffer from coherent rotation errors on the percent level, as the dominant error source in the experiment, and possibly spin imperfect readout \cite{ElbenXPlatform}. In the Supplementary Information, we argue, supported by detailed numerical simulations and previous work, that imperfect initial state preparation and measurement errors are well accounted for by a local depolarizing channel, i.e.~we modify our ansatz for EH in Eq.~(\ref{eq:method}) to 
\begin{equation}\label{eq:depol}
    \rho_A (\tilde{\boldsymbol{g}},p) \equiv (1-N_A p) \rho_A^{\text{coh}}(\tilde{\boldsymbol{g}}) + p \sum_{i \in A} \text{Tr}_i\left[\rho_A^{\text{coh}}(\tilde{\boldsymbol{g}})\right] \otimes \frac{\mathbb{1}_i}{2},
\end{equation} 
introducing a depolarization parameter $p$. In practice, we infer the parameter $p$ from the initial state at $t=0$, fitting the ansatz (\ref{eq:depol}) with $\rho_A^{\text{coh}} = \ket{\Psi_0}\bra{\Psi_0}$. At subsequent times we fix the parameter $p$ and perform EHT using the ansatz(\ref{eq:depol}) with $\rho_A^{\text{coh}}(\tilde{\boldsymbol{g}}) = \exp [-\tilde{H}_{A}(\tilde{\boldsymbol{g}})] /Z(\tilde{\boldsymbol{g}})$.
%($p=0.16$ in Fig.~(\ref{fig:Exp})).

While EHT is efficient in the number of measurements required, presently (\ref{eq:method}) is implemented as classical postprocessing of probabilities $P_U(\mathbf{s})$ measured on the quantum device, with corresponding requirements on classical computing, i.e.~feasible for subsystem sizes $N_A$ not much larger than $12$. Recent ideas \cite{Huang20, Bairey2019} allow, in principle, the determination of $\tilde{H}_A(\tilde{\boldsymbol{g}})$ in a scalable way, up to an unknown scaling factor for the coefficients $\tilde{\boldsymbol{g}}$. Determining the scaling factor, corresponding to the inverse temperature, opens interesting perspectives for measuring entanglement properties for larger subsystems. Variational quantum algorithms are potential candidates for extracting entanglement properties on a larger scale \cite{vqsd_coles, PhysRevA.101.062310}. In unpublished work we have developed hybrid classical-quantum algorithms where classical optimization of variational parameters is preceded by \textit{in situ} quantum postprocessing operations on $\rho_A$, with spins representing $A$ acting as quantum memory. 
\ \\

\section{Verification \& Fidelity Estimation} \label{Methods:XP}
We determine a (mixed-state) fidelity between the experimental quantum state under study, described by the density matrix $\rho_A \equiv \rho_1$, and  the reconstructed density matrix from EHT,  $\rho_A(\tilde{\boldsymbol{g}}) \equiv \rho_2$. To this end,  we consider the fidelity  \cite{Liang2019}
\begin{align}
\mathcal{F}_{\textrm{max}}(\rho_{1},\rho_{2})=\frac{\tr(\rho_{1}\rho_{2})}{\max\{\tr(\rho_{1}^{2}),\tr(\rho_{2}^{2})\}},
\label{eq:Fmax}
\end{align}
which measures the overlap between $\rho_{1}$ and $\rho_{2}$, respectively, normalized by their purities. As shown in Ref.~\cite{ElbenXPlatform}, $\mathcal{F}_{\textrm{max}}(\rho_{1},\rho_{2})$, i.e.\ terms of the form $\tr(\rho_i\rho_j)$ for $i,j=1,2$, can be evaluated from second-order \emph{cross-correlations} between the outcomes of randomised measurements. We set $P^{(1)}_U(\mathbf{s})$ the frequency of having observed a particular bitstring $\mathbf{s}$ in the experiment (where $\rho_A$ is realized) and $P^{(2)}_U(\mathbf{s})= \tr\left( \rho_A{(\boldsymbol{\tilde g}}) U\ket{\mathbf{s}}\bra{\mathbf{s}}U^\dagger \right)$. Then, we obtain  the overlap $\tr(\rho_i\rho_j)$  for $i=1,j=2$ and purities $\tr(\rho_i\rho_j)$ for $i=j=1,2$ via \cite{ElbenXPlatform}
\begin{align}
\tr(\rho_i\rho_j)= \label{eq:ovl} 
\frac{2^{N_{A}}}{N_U}\sum_U \sum_{\mathbf{s},\mathbf{s}'}(-2)^{-\mathcal{D}[\mathbf{s},\mathbf{s}']}{P_{U}^{(i)}(\mathbf{s})P_{U}^{(j)}(\mathbf{s}')},
\end{align}
where the Hamming distance $\mathcal{D}[\mathbf{s},\mathbf{s}']$ between two strings $\mathbf{s}$ and $\mathbf{s}'$ is defined as the number of local constituents  where $s_{k}\neq{s}'_{k}$, i.e.\ $\mathcal{D}[\mathbf{s},{\mathbf{s}}']\equiv\#\left\{ k\in \{1,\dots, N_A\}\,|\,s_{k}\neq{s}'_{k}\right\} $. 

Eq. (\ref{eq:ovl}) provides a direct experimental verification of the fidelity of the reconstructed density matrix, requiring no further theory input such as simulations, and can be evaluated from the same type of randomised measurements employed for EHT. Importantly, the measurements used for fidelity estimation should be independent from those used in EHT, to avoid false correlations and biasing. We note also that in principle more advanced measurement schemes exist, performing importance sampling on (\ref{eq:ovl}), thereby dramatically reducing the number of measurements required \cite{Flammia2011,da_Silva_2011}.
\ \\

\section{ Operator content of the EHT ansatz for quench experiments}\label{Methods:K}
In order to achieve good fidelities for EHT on quench experiments (see Fig.~\ref{fig:Exp}), the EHT ansatz $\tilde{H}_A(\boldsymbol{g})$ needs to be amended with additional operators $\mathcal{K}_A$ whose coefficients provide additional free fit parameters. CFT suggest lattice momenta, obtained from commutators of various terms of $H_A$, of the form
\begin{align}
\mathcal{K}_A^{(1)} &= \sum_{k<l \in A} \tilde{J}^{XY}_{kl}(\sigma^x_k\sigma^y_l - \sigma^y_k\sigma^x_l), \label{EqKA1}\\
\mathcal{K}^{(2)}_A &= \sum_{k<l}\sum_{m \neq k,l} \tilde{J}^{XYZ}_{klm}(\sigma^x_k\sigma^y_l\sigma^z_m - \sigma^y_k\sigma^x_l\sigma^z_m).\label{EqKA2}
\end{align}
Further improvements of the fidelity are obtained by including magnetization conserving operators of the form
\begin{align}
\mathcal{K}_A^{(3)} &= \sum_{k<l\in A} \tilde{J}_{kl}^{ZZ} \sigma^z_k\sigma^z_l + \! \! \sum_{k<l<m\in A}\! \! \! \tilde{J}^{ZZZ}_{klm} \sigma^z_k\sigma^z_l\sigma^z_m, \label{EqKA3}\\
\mathcal{K}^{(4)}_A &= \sum_{k<l}\sum_{m \neq k,l} \tilde{J}^{XXZ}_{klm}(\sigma^x_k\sigma^x_l\sigma^z_m +  \sigma^y_k\sigma^y_l\sigma^z_m).\label{EqKA4}
\end{align}

\section{ Quantum State Tomography} \label{Methods:QST}
For comparison with EHT, we consider in the main text two further well-known tomographical methods, Low Rank Least Squares (LRLS) and Projected Least Squares (PLS). Similar to EHT, these methods attempt to construct a density matrix estimator, $ {\rho}(X)$, depending on parameters $X$, by matching predictions of the estimator in a least squares sense with experimental observations from \textit{local} measurements performed on independent single copies of the density matrix. Here, the same cost function defined in Eq.~\ref{eq:method} is minimized, replacing $\tilde{\boldsymbol{g}}$ with $X$, with the only difference between LRLS and PLS being the form of the ansatz $\rho(X)$. 

\textit{Low-Rank Least Squares (LRLS)} \cite{Riofrio2017} takes as an ansatz $ \rho(X) = X^\dagger X$, i.e. matrices that are by construction positive semidefinite. Since the true density matrix is in practice often not full rank, $X$ can be a rectangular (complex valued) $r \times d$ matrix, with $r$ the rank of the density matrix estimator. The total measurement effort needed to obtain a fixed precision scales proportional to the number of unknown variables, i.e. the $r \times d$ entries of $X$. This method comes with a substantial computational overhead associated with determining the $r \times d$ unknown variables.

\textit{Projected Least Squares (PLS)} \cite{Sugiyama_2013, Guta20} consists of first finding the Hermitian, but not necessarily positive semidefinite, matrix that would produce the observations exactly. This can be done analytically, thereby eliminating the computational overhead of LRLS. For a given set of measurement results in the computational basis, obtained after applying basis transformations U, the resulting matrix is given by
\begin{align}\label{EqRhoRT}
 {\rho}_{RT} = \sum_{\mathbf{s}, \mathbf{s}'} \sum_{ {U}}P_U(\mathbf{s}) (-2)^{-\mathcal{D}[\mathbf{s}, \mathbf{s}']}  {U} \ket{\mathbf{s}'} \bra{\mathbf{s}'}  U^{\dagger},
\end{align}
with $\mathcal{D}[\mathbf{s},\mathbf{s}']$ the Hamming distance defined in Eq.~(\ref{eq:ovl}).
For the randomised local unitary transformations $ {U}$ considered here, the estimator $ {\rho}_{RT}$ is the one obtained in randomised tomography \cite{Elben19PRA} and shadow tomography \cite{Huang20}.

The matrix (\ref{EqRhoRT}) is generally not positive semi-definite, and hence does not represent a physical density matrix. However, the matrix can be projected onto the space of positive semi-definite matrices via a simple procedure, by rescaling the eigenvalues and truncating them to positive values \cite{Smolin12}. 
Again, the scaling of the number of measurements needed to obtain a certain fidelity is exponential in the system size and proportional to $r \times d$, where $r$ is a measure of the effective rank of the reduced density matrix.

\section{Entanglement Hamiltonian Tomography in the Presence of Decoherence and Imperfections} \label{app:decoh}

% In the following we provide a detailed analysis of EHT in the presence of decoherence and measurement errors in the . We perform EHT on  quantum states obtained from numerically simulating unitary quench dynamics and including decoherence models and experimental error rates of Ref.~\cite{Brydges2019} (see also below). In particular we show, by taking into account experimental error rates that decoherence during time evolution is weak, and we show explicitly that EHT can be applied in this case.  Furthermore, we show that  measurement errors, from local decoherence and miscalibration of random unitaries on the quantum hardware, can be mitigated  by modifying the ansatz for the reduced density matrix $\rho_A(\tilde{\boldsymbol{g}})$.

Entanglement Hamiltonian Tomography (EHT) in quench dynamics is based on an ansatz for the EH involving quasi-local few-body terms. A priori, theoretical considerations support such an ansatz for subsystem density matrices arising in closed system dynamics. However, for the experiment discussed around Fig.~1 in the main text, we have to deal with various sources of imperfections. The most important are $\textit{i})$ Decoherence during time evolution, i.e. spin-flip errors and spontaneous emission; \textit{ii}) Imperfect preparation of the initial state; \textit{iii}) Calibration errors of the random unitaries applied on the quantum hardware. 

Below we provide a detailed analysis of EHT including decoherence and measurement errors, in the context of the experiment described in the main text. In particular we show, by taking into account experimental error rates, that decoherence during time evolution is negligible, and that the quality of the density matrix extracted from EHT is only marginally affected. Furthermore, we show that the miscalibration of random unitaries on the quantum hardware, can be mitigated  by modifying the ansatz for the reduced density. This is achieved with the ansatz \begin{equation}\label{eq:locnoiseansatz}
    \rho_A(\tilde{\boldsymbol{g}}) = (1-pN)\rho_A^\text{coh}(\tilde{\boldsymbol{g}})+ \sum_i \text{Tr}_i(\rho_A^\text{coh}(\tilde{\boldsymbol{g}})) \otimes \mathbb{1}_i,
\end{equation} with $p$ a parameter describing local depolarising noise. Fitting with such an ansatz allows to extract the coherent part $\rho_A^\text{coh}(\tilde{\boldsymbol{g}})$ and thus to partially mitigate measurement errors. Below we discuss these error sources one by one, providing justification for the above noise model in context of the experiment.

%In particular we show, by taking into account experimental error rates, that decoherence during time evolution is weak and thus causes only a slight deviation from the reduced density matrix $\rho_A$ in coherent dynamics. We demonstrate that our tomographic protocol can still be applied in this case. This is because dissipation typically suppresses off-diagonal elements in $\rho_A$, causing a simple structure of the reduced density matrix which can be described by a quasi-local EH ansatz. 

\textit{Decoherence during time evolution --}
We numerically simulate the dynamics for an ion string of $N=10$ sites, governed by the Hamiltonian
\begin{align} \label{eq:xy}
    \frac{H_{XY}}{\hbar} = \sum_{i,,<i} J_{ij} \left( \sigma_i^+ \sigma_j^- + \text{H.c.} \right) + B \sum_i \sigma_i^z.
\end{align}
with $J_{ij} \sim 1/|i-j|^{\gamma}$ and $\gamma \approx 1.24$.  
The initial state is modeled as
\begin{align}
    \rho_0 = \bigotimes_{i=1}^N 
    \begin{pmatrix}
 p_i & 0\\
 0 & 1-p_i
\end{pmatrix}
\end{align}
with $p_i = 0.004$ for $i$ even and $p_i = 0.995$ for $i$ odd, resulting in a total initial purity of $\text{Tr}(\rho_0^2) \approx 0.91$. The time evolution is calculated by numerically integrating the Lindblad master equation
\begin{align} \label{eq:lindblad}
    \begin{split}
        \dot{\rho}(t) &= -\frac{i}{\hbar} \left[ H_{XY}, \rho(t) \right] \\
        &+ \sum_i \frac{1}{2} \left[ 2 C_i \rho(t) C_i^{\dagger} - \rho(t) C_i^{\dagger} C_i - C_i^{\dagger} C_i \rho(t) \right].
    \end{split}
\end{align}
Here the jump operators describe local spin-flips $C_i = \sqrt{\gamma_F} \sigma_i^x$ (for $i=1\dots N$) and spontaneous decay of the ions $C_{i+N} = \sqrt{\gamma_D} \sigma_i^-$ (for $i=1\dots N$) respectively. The simulation is performed using the experimental decay rates $\gamma_D \approx \gamma_F \approx 0.7\,s^{-1}$. To a good approximation the dynamics is constrained to a subspace of constant total magnetization $S^z_{\text{tot}} = \sum_i \sigma_i^z$, which is a decoherence-free subspace with respect to global dephasing.  

\begin{figure}[t]
	\centering
	\includegraphics[width=0.97\linewidth]{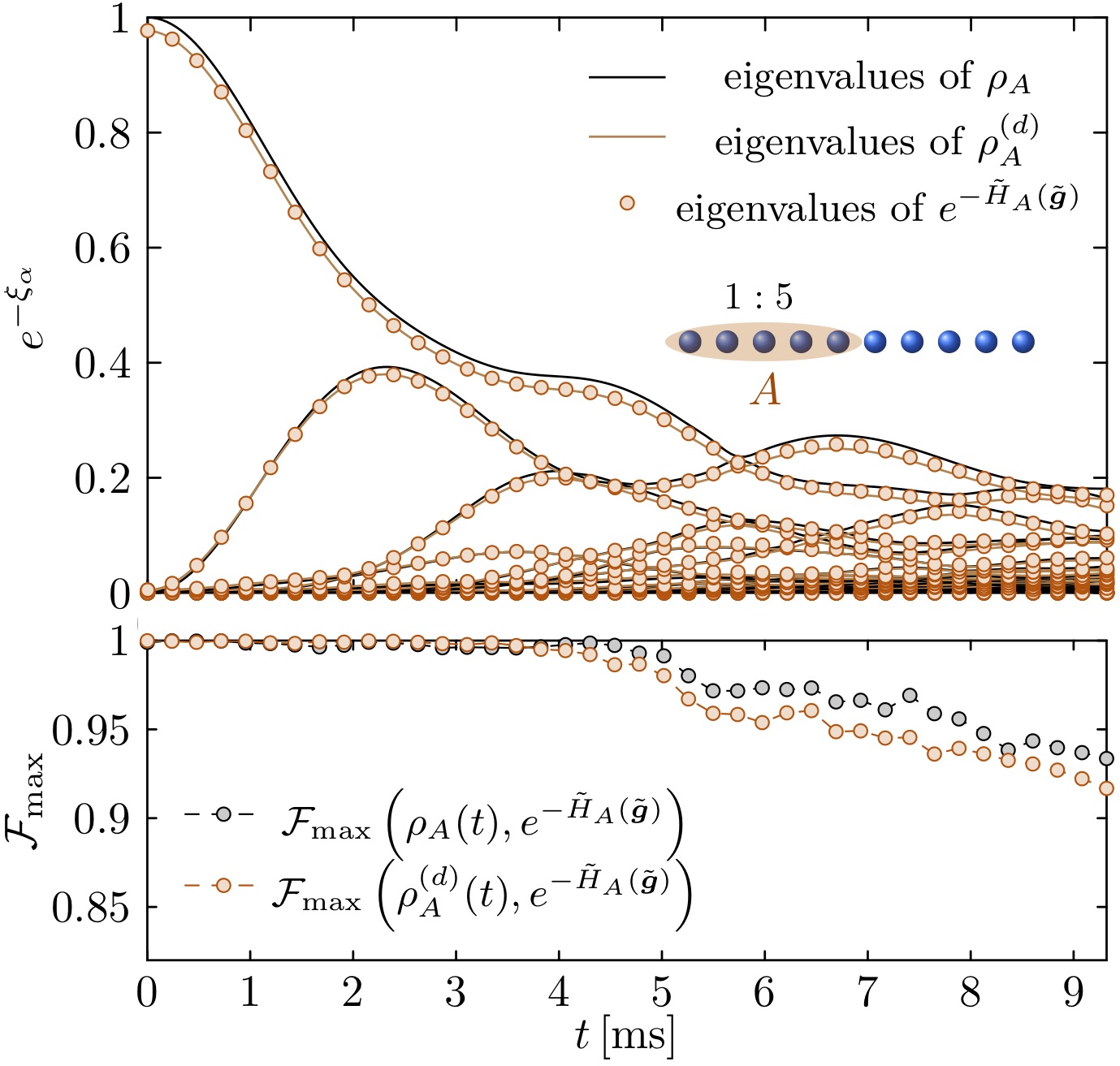}
	\caption{\textit{Theoretical simulation of EHT for quench dynamics in the long-range XY-model including local decoherence.} a) Schmidt decomposition as a function of time during coherent and dissipative dynamics for a subsystem of $N_A=5$ shown as an inset. Orange points represent the eigenvalues of the ansatz $\rho_A(\tilde{\boldsymbol{g}})$ fitted to the density matrices $\rho_A^{(d)}(t) = \text{Tr}_{\bar{A}}[\rho(t)]$ with $\rho(t)$ obtained from (\ref{eq:lindblad}). b) Maximum fidelities for an ansatz $e^{-\tilde{H}_A(\tilde{\boldsymbol{g}})}$ fitted on $\rho_A(t)$ (obtained from coherent dynamics) with respect to $\rho_A(t)$, and for the ansatz $\rho_A(\tilde{\boldsymbol{g}})$ (see main text) fitted on $\rho_A^{(d)}(t)$ with respect to $\rho_A^{(d)}(t)$.  }
	\label{fig:QuenchDecoherence}
\end{figure}

Fig.~\ref{fig:QuenchDecoherence} a) shows the Schmidt decomposition for a subsystem of $N_A = 5$ sites at the boundary as a function of time, resulting from coherent time evolution, as well as from integrating the master equation (\ref{eq:lindblad}). Clearly, the decoherence has only a small effect on the Schmidt values, and does not vary significantly over time. We fit an ansatz of the form
$\rho_A(\tilde{\boldsymbol{g}}) = e^{-\tilde{H}_A(\tilde{\boldsymbol{g}}) }$ to the reduced density matrices $\rho_A^{(d)}(t) = \text{Tr}_{\bar{A}} [\rho(t)]$ extracted from states $\rho(t)$, which include dissipation. The ansatz for the EH is equivalent to the ansatz described in the main text, i.e.\ contains 3-body terms including $\mathcal{K}_A^{(3)}$ and $\mathcal{K}_A^{(4)}$ (see Appendix \ref{Methods:K}). For the analysis described here, we adopt the Frobenius norm of the matrix difference $|\!| \rho_A^{d}(t) - \rho_A(\tilde{\boldsymbol{g}}) |\!|$ as a cost function for the fitting. As shown in Fig.~\ref{fig:QuenchDecoherence} a), the Schmidt decomposition obtained from the ansatz $\rho_A(\tilde{\boldsymbol{g}})$ coincides well with the eigenvalues of $\rho_A^{(d)}$. Fig.~\ref{fig:QuenchDecoherence} b) demonstrates the performance of EHT on states from coherent and dissipative dynamics, by showing the maximum fidelity with respect to the exact reduced density matrices as a function of time. Fig.~\ref{fig:QuenchDecoherence} shows that the ansatz is capable of describing dissipative dynamics, accompanied by a small drop in fidelity of the order of 1-2\%.   We attribute this to the fact that dissipation typically suppresses off-diagonal elements in $\rho_A$, causing a simple structure of the reduced density matrix which can be described by a quasi-local EH ansatz.

\textit{-- Measurement errors.} Measurement errors in EHT predominately arise from local decoherence (local depolarisation) during the application of the local random unitaries, and due to unitary errors in the realization of the random unitaries caused by small miscalibrations of the quantum hardware. As analysed in detail in Ref.~\cite{Elben_2020},   such miscalibration effects can be modeled by assuming that, instead of a unitary $U_A$, the device implements a unitary $W_A = U_A V_A$ with $V_A = \bigotimes_{j=1}^{N_A} \exp\left[{i h_j(\nu)}\right]$. Here $h_j(\nu)$ are random hermitian matrices for the particles $i \in A$, where the real and imaginary part of each matrix element is independently distributed according to the standard normal distribution with mean value zero standard deviation $\nu$. Thus $\nu$ quantifies the value of miscalibration, with $\nu = 0$ corresponding to a perfect match between the unitaries. Averaging over $V_A$, one can show \cite{Elben_2020}, that such miscalibration effectively acts as local depolarisation noise.

%Numerical simulations show that such a miscalibration between the unitaries causes the density matrix obtained from EHT to get more mixed. 
%As shown in Ref.~\cite{Elben_2020}, such miscalibration , modeled as local depolarising noise.
We can thus try to correct for effects of local decoherence and miscalibration, by modifying the EHT ansatz with a  local depolarising channel with a variational parameter $p$.
%\begin{align} \label{eq:locnoiseansatz} \rho_A(\tilde{\boldsymbol{g}}) = (1 \!-\! N_A p) e^{-\tilde{H}_A(\tilde{\boldsymbol{g}})} + \!p \! \sum_{i \in A} \text{Tr}_i \! \left( e^{-\tilde{H}_A(\tilde{\boldsymbol{g}})} \right) \otimes \frac{\mathbb{1}_i}{2}.
%\end{align}
Fitting Eq.~\ref{eq:locnoiseansatz} to experimentally observed frequencies allows then to extract $e^{-\tilde{H}_A(\tilde{\boldsymbol{g}})}$ from the ansatz. 
\begin{figure}[t]
	\centering
	\includegraphics[width=0.97\linewidth]{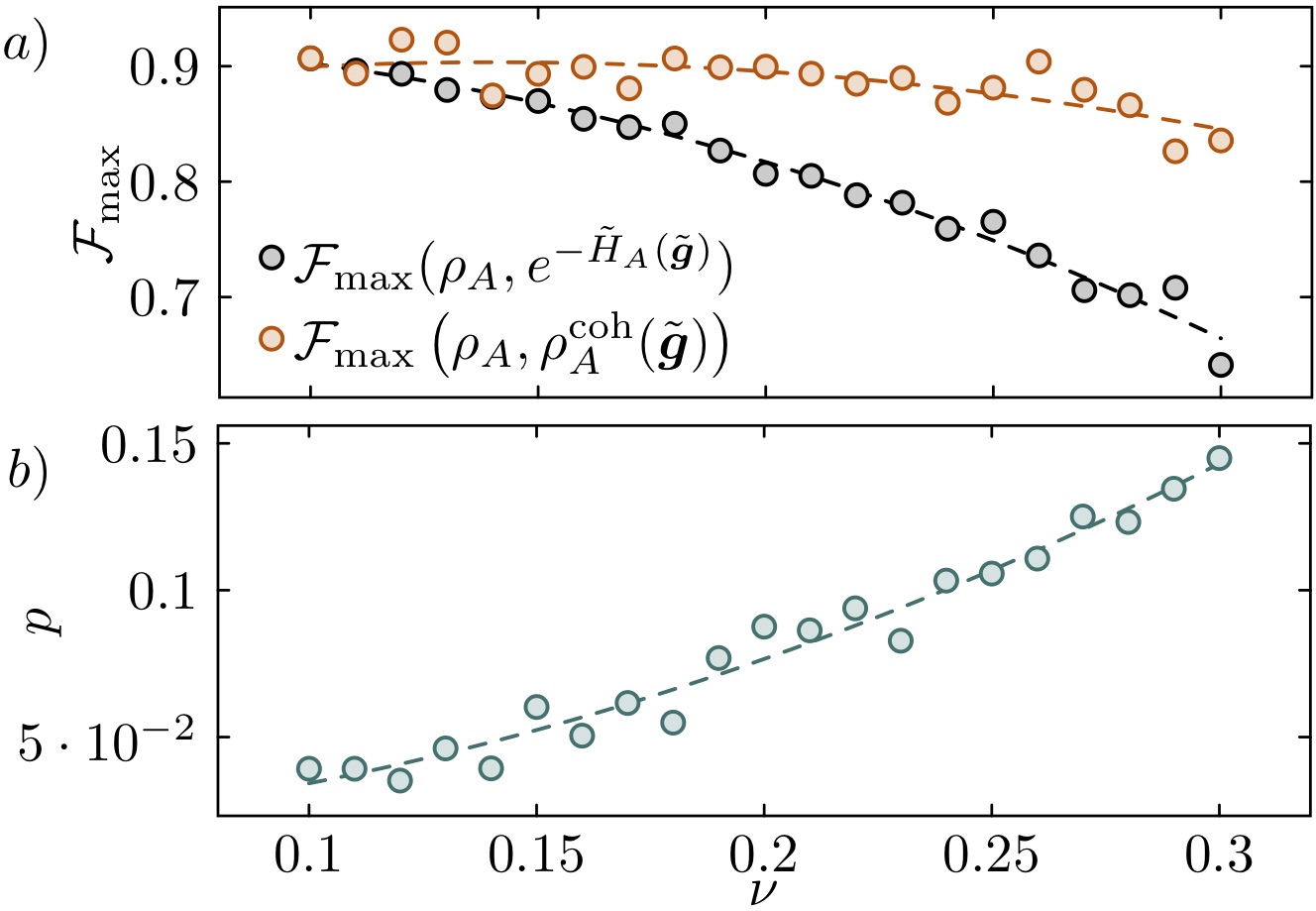}
	\caption{\textit{EHT with a modified EH ansatz on miscalibrated random measurements.} a) Maximum fidelities with respect to the exact density matrix $\rho_A$ obtained from simulation of coherent dynamics. Black points show the fidelity obtained by fitting with an ansatz of the form $e^{-\tilde{H}_A(\tilde{\boldsymbol{g}})}$. For the orange points EHT is performed using the ansatz Eq.~(\ref{eq:locnoiseansatz}) and the coherent part is extracted from the result via $\rho_A^{\text{coh}}(\tilde{\boldsymbol{g}}) = \rho_A(\tilde{\boldsymbol{g}}) (1-N_Ap)^{-1} - p (1-N_Ap)^{-1} \sum_{i \in A} \text{Tr}_i \! \left( e^{-\tilde{H}_A(\tilde{\boldsymbol{g}})} \right) \otimes \mathbb{1}_i/2$. The fidelity is calculated with respect to $\rho_A^{\text{coh}}(\tilde{\boldsymbol{g}})$. }
	\label{fig:miscalibration}
\end{figure}

In the following, we test this approach on a representative state, obtained from numerically simulating coherent dynamics with the Hamiltonian (\ref{eq:xy}) for 10 sites. We simulate EHT for $N_A = 5$ sites on the boundary, sampling the reduced density matrix with the unitaries $W_A$ while frequencies from the EH ansatz are obtained using the unitaries $U_A$. To mimic the experimental situation, we choose $N_U = 500$ unitaries and $N_M = 150$ samples per unitary. 

Fig.~\ref{fig:miscalibration} a) compares the maximum fidelities obtained by the standard procedure of fitting the ansatz $e^{-\tilde{H}_A(\tilde{\boldsymbol{g}})}$ to the data, and by optimizing the ansatz (\ref{eq:locnoiseansatz}) with subsequent extraction of the $e^{-\tilde{H}_A(\tilde{\boldsymbol{g}})}$-part as function of the level of miscalibration $\nu$. The fidelity is calculated with respect to the theoretically exact reduced density matrix, obtained from simulating coherent dynamics. Fig.~\ref{fig:miscalibration} b) shows the rising trend of the $p$-parameter obtained from the fitting as the level of miscalibration increases. As demonstrated in Fig.~\ref{fig:miscalibration}, this procedure allows to partially filter out calibration errors of the random unitaries, keeping the maximum fidelity on a high level as a function of miscalibration $\nu$.

\section{EHT for Gibbs states}

Here we investigate EHT for  Gibbs states $\rho_A = e^{-\beta H_A}$ for a Hamiltonian $H_A$ and  inverse temperature $\beta$.  In particular,  we perform a scaling analysis of the required number of experimental runs with system size, for different temperatures. Our numerical results demonstrate that EHT provides a significant advantage in terms of the required number of measurements compared to PLS tomography. This holds  in particular  for Gibbs states at high temperatures (small $\beta$) where  $\rho_A$ has high rank, and thus tomographic methods based on low-rank assumptions are not efficiently applicable.    

\begin{figure}[t]
	\centering
	\includegraphics[width=0.97\linewidth]{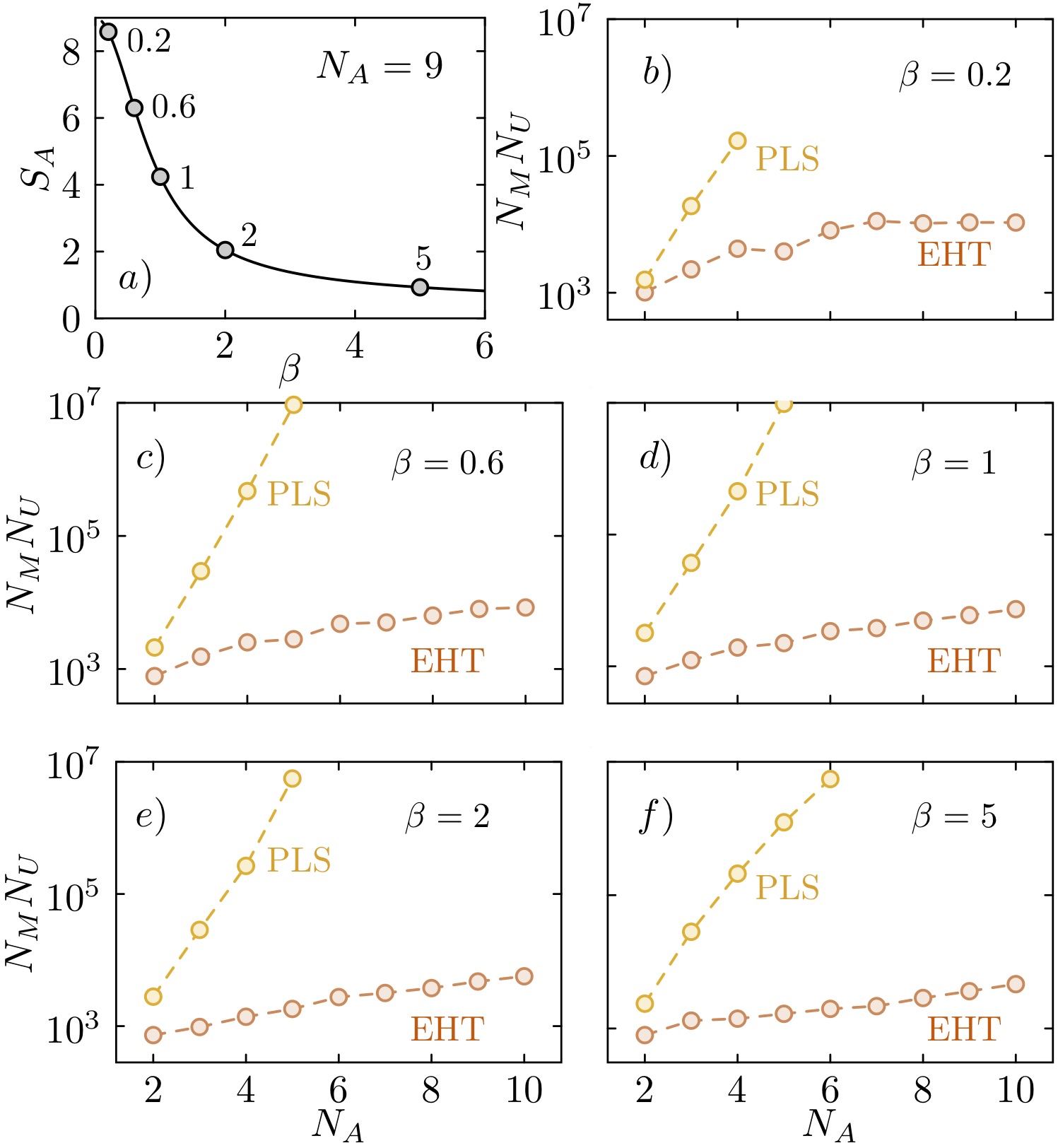}
	\caption{\textit{Scaling analysis for EHT on Gibbs states} a) Von Neumann entropy for thermal states of the transverse-field Ising model as a function of $\beta$ for a subsystem of $N_A = 9$ sites. b)-f) scaling analysis for PLS and EHT with the EH ansatz given in the main text. The panels show the number of measurements $N_UN_M$ required to reach an infidelity of $\epsilon=10^-2$ with respect to the exact density matrix $\rho_A$, as a function of the subsystem size $N_A$. For EHT $N_U=2$ and for PLS $N_M=1$ are fixed.}
	\label{fig:gibbs}
\end{figure}

To be specific, we consider Gibbs states $\rho_A = e^{-\beta H_A}$ for the transverse-field Ising model 
\begin{equation}
    H_A = J\sum_{i \in A} \sigma_i^x \sigma_{i+1}^x + B \sum_{i \in A} \sigma_i^z
\end{equation} with $J=1$ and $B=0.9$ as the underlying Hamiltonian.  In Fig.~\ref{fig:gibbs} a), we display the Von Neumann entropy of $\rho_A$ for a system size of $N_A=9$ as function of the inverse temperature $\beta$, showing that small $\beta$ correspond to highly mixed states.  

To perform EHT on $\rho_A$, we make an ansatz 
\begin{equation}
    \rho_A(\tilde{\boldsymbol{g}})=\exp\left( -\tilde{H}_A(\tilde{\boldsymbol{g}})\right)
\end{equation} with the EH $\tilde{H}_A(\tilde{\boldsymbol{g}})$
 as a deformation of the Ising Hamiltonian 
 \begin{equation}
    \tilde{H}_A(\tilde{\boldsymbol{g}}) = \sum_{i \in A} \tilde{J}_i \sigma_i^x \sigma_{i+1}^x +  \sum_{i \in A} \tilde{B}_i \sigma_i^z,
\end{equation} with free fit parameters $\tilde{J}_i$ and $\tilde{B}_i$. We numerically simulate  $N_U N_M$ randomized measurements  on $\rho_A$ and fit $\rho_A(\tilde{\boldsymbol{g}})$ to the outcomes, according to Eq.~(A1).  

In Fig.~\ref{fig:gibbs}  panels b-d), the number of required randomized measurements $N_UN_M$ is plotted to achieve an Uhlmann infidelity  of $\rho_A(\tilde{\boldsymbol{g}})$ to $\rho_A$ below the threshold of $10^{-2}$.  Additionally, we show the required number of measurements for quantum state tomography using projected least squares (PLS) (see Appendix \ref{Methods:QST}). Our analysis is compatible with a sub-exponential scaling of EHT, and consistent with rigorous results from Hamiltonian learning on Gibbs states \cite{anshu2020sampleefficient}.  In particular, this scaling persists for all shown temperatures, with an increasing absolute number of measurements  with raising temperature,  by approximately one order of magnitude from $\beta = 5$ to $\beta=0.2$.  

\section{Conformal Field Theory \& Entanglement Hamiltonian} \label{supp:CFT}

Here we provide a brief discussion of Conformal Field Theory and Entanglement Hamiltonians as background for discussions in the main text.

Conformal field theory (CFT) in $(1+1)$ dimensions provides us with explicit expression for entanglement (modular) Hamiltonians \cite{Cardy2016,Wen2018}. 
For connected subsystems $A$, it has been shown that for a broad range of CFT states, including ground states, thermal states and states generated via quench dynamics,  the entanglement Hamiltonian density $H_E(t,x)$ can be written in terms of the energy   momentum tensor $T(x,t)$  multiplied with a local weight factors \cite{Cardy2016,Wen2018}. This suggests ans\"atze  for the EH also for the lattice models considered in this work, which, complemented with additional few body terms, provide an accurate description of the EH, even beyond the regime of applicability of a CFT description (see e.g.~Fig.~1 of main text). 

In the following, we  briefly summarize CFT predictions for entanglement Hamiltonians. We take $A=[0,l]$ to be a partition at the end of a system $S=[0,L]$, with $L \gg l$ much larger than any length scale in the system, such that the right boundary $L$ can be neglected. We consider a quantum quench from an initial state $|\psi_0\rangle \sim  e^{-(\beta_0/4) H_{\text{CFT}}}|b\rangle$ where $b\rangle$ is a conformally invariant boundary state and $\beta_0 > 0$ introduces a finite correlation length $\ell \sim \beta_0$ \cite{Calabrese2016,Cardy2016,Wen2018}. It has been shown that in this setting, $|\psi_0\rangle$  represents generic ground states of  Hamiltonians $H_0$ with inverse mass gap $m_0^{-1}\sim \beta_0$ \cite{Calabrese2016}. 
For this initial state,  the entanglement Hamiltonian 
\begin{align}
    e^{-H_E(t)} = \tr_A\left[ e^{-i H_{\textrm{CFT}}t} \ket{\psi_0}\bra{\psi_0 }e^{i H_{\textrm{CFT}}t} \right] 
\end{align}
 can be calculated for all times $t$  exactly via a conformal mapping to an annulus. While we refer for the general result to Refs.~\cite{Cardy2016,Wen2018}, we consider here illustrative limiting cases. 
 
 \paragraph{Ground state with long-range correlations}
 For $\beta_0 \gg l$,  $|\psi_0\rangle$ is locally indistinguishable from the ground state of the CFT $  \ket{ \psi_{\textrm{GS}}}$. The partition size $l$ is the only remaining length-scale and 
the entanglement Hamiltonian  is given by the Hamiltonian density $\mathcal{H}(x)$ modified with a parabolic weight factor \cite{hislop1982,Cardy2016,Wen2018,casini2011}
  \begin{align}
     \left. H_E(0) \right|_{\beta_0 \gg l} \simeq	 \int_0^l \! \text{d}x  \, \frac{l^2-x^2}{2l} \mathcal{H}(x) \; .
\end{align}
We note that this is a direct CFT generalization of the  Bisognano-Wichmann theorem to describe the entanglement Hamiltonian in a finite interval $A=[0,l]$ embedded in a semi-infinite system \cite{Cardy2016,Wen2018}. As expected,  close to the entanglement cut at $x=l$, the local weight factor increases linearly $\sim (l-x)$.

\paragraph{Ground state with short-range correlations}
For $\beta_0 \ll l$, $|\psi_0\rangle$ represents the groundstate of a generic many-body Hamiltonian with (short) correlation length $\beta_0$ \cite{Calabrese2016,Cardy2016,Wen2018}. The entanglement Hamiltonian  is given by \cite{Cardy2016,Wen2018}
 \begin{align}
    \left. H_E(0)\right|_{\beta_0 \ll  l }  \simeq	 {\beta_0} \int_0^l \text{d}x \sinh\left[ \frac{2\pi}{\beta_0}(l-x)\right] \mathcal{H}(x) \;.
\end{align}
For a short range entangled state, the dominant contribution to the Schmidt spectrum (i.e.\ to the entanglement between $A$ and the remainder of the system) is expected to arise from regions close to the entanglement cut $(l-x) \ll \beta_0$. Here, the hyperbolic weight factor can be expanded to linear order $\sim (l-x)$, as expected from the Bisognano-Wichmann theorem. Contributions from far from the boundary, where the entanglement Hamiltonian density is large, are exponentially suppressed.

\paragraph{Quantum quench}
For general $t>0$, the entanglement Hamiltonian receives contributions from  energy  $T_{00}(x)=\mathcal{H}(x)$ and momentum $T_{01}(x)$ density \cite{Cardy2016,Wen2018}. The latter can be interpreted as emerging quasi-particle currents spreading entanglement through the system \cite{Zhu2020} and can be represented as additional few body (momentum) terms in the EH (see main text and App.~\ref{sec:criticality}).
At long times $t \gg l \gg \beta_0 $, when the system approaches thermal equilibrium, the expressions considerably simplify. The contributions from the momentum density vanish and the entanglement Hamiltonian $
    H_E(t\rightarrow \infty)  \simeq	 \int_0^l \! \text{d}x  \, \beta(x) \, \mathcal{H}(x)
$ is  determined by  \cite{Cardy2016,Wen2018}
\begin{align}
   \beta(x) = 2 \beta_0 \, \frac{\sinh\left[  \pi (l-x) / \beta_0 \right]\sinh\left[  \pi (l+x) / \beta_0 \right]}{\sinh\left[ 2 \pi l / \beta_0 \right]}
\end{align}
Remarkably,  $ H_E(t\rightarrow \infty)  $ equals \textit{exactly} to the entanglement Hamiltonian in the thermal state \cite{Cardy2016,Wen2018}
\begin{align}\exp[-H_E(t\rightarrow \infty)]= \tr_A\left[ \exp(-\beta_0 H_{\text{CFT}}) \right] .
\end{align}
Thus, CFT provides an explicit demonstration of the eigenstate thermalization hypothesis \cite{Deutsch1991,Srednicki1994,Rigol2008,Garrison2018}. The weight factor $\beta(x)$ can be  interpreted as a spatially varying (inverse) temperature. Close to the edge $x\sim l$, $\beta(x)^{-1}$
is largest as entropy is generated at the entanglement cut. With the distance from the cut, the temperature decreases, and saturates to the thermal value $\beta(0)^{-1}=\beta_0^{-1}$.

\section{Ansatz extensions beyond CFT}
\label{sec:criticality}
In the analysis of a global quench in the quantum Ising model near criticality (see main text), we discuss the inclusion of lattice momentum terms in the EH ansatz, describing propagation of quasi particles spreading entanglement. It is noted that such terms lead to a significant enhancement of the ansatz at early times after the quench. In the following we quantify the role of such terms by analysing the density matrix fidelities with respect to the exact $\rho_A$ as a function of time. In particular we demonstrate that terms corresponding to lattice analogs of the momentum density are the only type of 2-body terms which yield an enhancement in approximating the theoretically exact density matrix $\rho_A$.  This is consistent with and provides a testbed for the validity of  effective CFT predictions in lattice models (see also main text and App.~\ref{supp:CFT}).

\begin{figure}[t]
	\centering
	\includegraphics[width=0.97\linewidth]{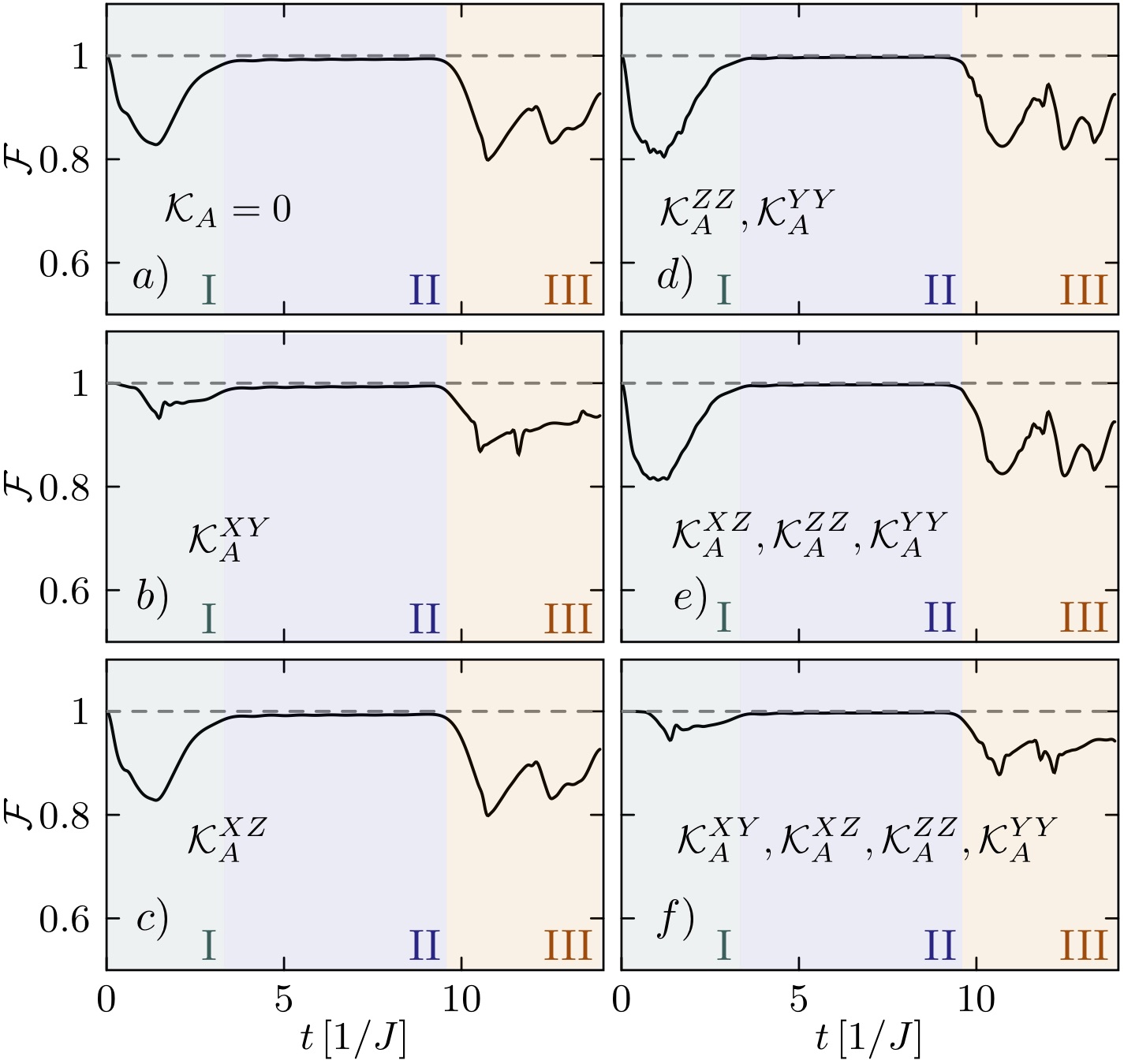}
	\caption{\textit{Density matrix fidelities for differnt EH ansatz extensions in a global quench to the critical point of the quantum Ising model.} Uhlmann fidelities of the ansatz $\rho_A(\tilde{\boldsymbol{g}}) = e^{-\tilde{H}_A(\tilde{\boldsymbol{g}})}$ with $\tilde{H}_A(\tilde{\boldsymbol{g}}) = \sum_{i} \tilde{J}_{i,i+1} \sigma_i^x \sigma_{i+1}^x + \sum_i \tilde{B}_i \sigma_i^z + \mathcal{K}_A$ with respect to the theoretically exact density matrix $\rho_A$ as afunction of time. The extensions $\mathcal{K}_A$ consist of 2-body Pauli terms as depicted in the insets. }
	\label{fig:momentumcontributions}
\end{figure}
In addition to the situation described in the main text, we include 2-body Pauli terms into the EH ansatz $\mathcal{K}_A^{\alpha, \beta} = \sum_{i,j} \sigma_i^{\alpha} \sigma_j^{\beta}$, with $\alpha, \beta = \{ x, y, z \}$  denoting the Cartesian coordinates. Fig.~\ref{fig:momentumcontributions} shows the fidelities with respect to the exact density matrices $\rho_A(t)$ for different ansatz extensions $\mathcal{K}_A$ as a function of time. While adding $XY$-terms, which correspond to lattice momentum contributions, yield a fidelity enhancement in region I [panels b) and f)], other types of 2-body terms [panels (c-e)] do not improve the achievable fidelity. In region III, where quasi particles propagate back into the subsystem, 2-body terms do not suffice to push the fidelity above the 90\% threshold. In this region additional 3-body terms have to be included in the ansatz.  

\bibliographystyle{apsrev4-1}    
%\bibliography{lit}{}

\end{document}